\documentclass{aa}  

\usepackage{graphicx}
\usepackage{txfonts}
\usepackage{caption}
\usepackage{subcaption}
\usepackage{footnote}
\usepackage{longtable}
\usepackage{array}
\usepackage[symbol]{footmisc}

\begin{document}

   \title{
   Color measurements of the polarized light scattered by \\
   the dust in protoplanetary disks}
   \author{J. Ma, \inst{1}
   H.M. Schmid, \inst{1}
   T. Stolker, \inst{2}
          }

   \institute{Institute for Particle Physics and Astrophysics, 
          ETH Zurich, Wolfgang Pauli Strasse 17, CH-8093 Zurich\\
              \email{jma@phys.ethz.ch}
              \and 
              Leiden Observatory, Leiden University, Niels Bohrweg 2, 2333 CA Leiden, The Netherlands}
              
   \date{Received --; accepted --}

  \abstract
   {Many reflected light images of protoplanetary disks have been obtained with ground-based high-contrast instruments. Quantitative measurements of the reflected radiation provide strong constraints on the scattering dust which can clarify the dust particle evolution in these disks and the composition of the forming
   planets. }
   {We derived the wavelength dependence of the polarized reflectivity $(\hat{Q}_{\varphi}/I_\star)_\lambda$ or color for a sample of disks to contrain the dust based on these data. Further we searched for systematic differences in the dust properties between subgroups of disks.}
   {We selected from the ESO archive polarized imaging for 11 protoplanetary disks obtained at visible and near-infrared wavelengths with the SPHERE/ZIMPOL and SPHERE/IRDIS high contrast instruments. All disks have bright and well-resolved structures, such as rings or spirals, which allow accurate determinations of the intrinsic polarized reflectivity $\hat{Q}_\varphi/I_\star$ at multiple wavelengths. For this, we corrected the observations for the smearing effects caused by the point spread functions (PSFs) of the individual dataset with a novel correction procedure applicable to inclined disks. For the 11 disks, we derived a total of 31 $\hat{Q}_{\varphi}/I_{\star}$ values for wavelengths from 0.62~$\mu$m ($R$ band) to 2.2~$\mu$m ($Ks$ band) and compared our results, if possible, with previous determinations.
   For each disk, we derived a polarized reflectivity color $\eta_{V/IR}$ between a visible band $\lambda<1~\mu$m and a near-IR band $\lambda>1~\mu$m and other wavelength combinations.  
   We also consider model calculations for the polarized reflectivity colors $\eta$ for protoplanetary disks to constrain the scattering properties of the dust. 
   }
   {The polarized reflectivities for the investigated disks structures are at a typical level between $Q_\varphi/I_\star\approx 0.1$~\% to 1.0~\%. These values depend strongly on the observing conditions and a correction for the PSF smearing effects is essential to obtain the intrinsic values $\hat{Q}_\varphi/I_\star$. Corrected values $\hat{Q}_\varphi/I_\star$ are on average about a factor of 1.6 higher than the observed values. We checked the accuracy of the PSF calibrations procedure with simulations and literature data, and show that the large systematic errors in the observational values are reduced to a relative level $\Delta\hat{Q}_\varphi/\hat{Q}_\varphi\approx 10~\%$ or even less. 
   The high accuracy yields wavelength gradients for the polarized reflectivity $(\hat{Q}_\varphi/I_\star)_\lambda$ or colors $\eta$ which are significantly different between different objects. We find in our sample for all disks around Herbig stars (HD~169142, HD~135334B, HD~100453, MWC~758, and HD~142527) a red color $\eta_{\rm V/IR}>0.5$, while four out of six disks around T-Tauri stars (PDS~70, TW~Hya, RX~J1615, and PDS~66) are gray $-0.5<\eta_{\rm V/IR}<0.5$.
   The red colors support the presence of rather compact dust grains, while the absence of blue colors (except for the near-infrared color of PDS~66) is not compatible with very porous aggregates composed of small monomers. We suspect, that the very red colors $\eta_{\rm V/IR}\approx 2$ obtained for LkCa~15 and MWC~758 could be the result of an "extra" reddening of the radiation illuminating the disk caused by absorbing hot dust near the star. 
   We discuss the prospects of further improvements for the derivation of dust properties in these disks if the fractional polarization $\langle p_\varphi \rangle$ or other parameters of the reflected light are also taken into account in future studies.
   }
   {}
   

   \maketitle
%

\section{Introduction}
The dust particles in protoplanetary disks undergo significant evolution in their size distribution, structure, and composition. This changes the hydrodynamical properties of the dust and their location and motion in the disk. 
It is important to understand these complex processes better because dust plays an essential role in the planet formation within these disks. 

The dust is illuminated by stellar radiation, absorbs, and also scatters the stellar light. The heated dust produces thermal infrared (IR) emissions which provide information about the dust temperature and the amount of radiation energy absorbed at different separations from the star and indicate dust composition from mineral emission features \citep[e.g.,][]{Woitke2016}. The stellar light scattered by the dust allows for the determination of the disk geometry in reflected light from high-resolution observations \citep[e.g.,][]{Benisty2022}. 
The analysis of the reflected light provides valuable insights into the dust scattering parameters. The goal of this work is to extract the reflected polarized light for characterizing the dust properties.
Obtaining and analyzing such data for many circumstellar disks is beneficial in finding systematic dependencies of dust properties as a function of system parameters, which can constrain the dust evolution processes occurring in planet-forming disks.

Measuring the scattered radiation from protoplanetary disks is challenging because the faint disk signal, which typically has an apparent radius less than one arcsec, needs to be disentangled from the much stronger signal of the central star. 
Successful disk observations were achieved with the \textit{Hubble} Space Telescope (HST) \citep[e.g.,][]{Silber2000, Grady2001, Krist2005, Perrin2009, Debes2013} thanks to its good spatial resolution and stable point spread function (PSF). Many protoplanetary disks have also been observed with ground-based telescopes using adaptive optics (AO) systems and imaging polarimetry \citep[e.g.,][]{Benisty2022, Fukagawa2010, Avenhaus2018, Garufi2020, Monnier2019}.
This technique is particularly beneficial because the scattered and therefore polarized light from the disk can be distinguished from the much brighter, direct, and therefore unpolarized light from the star, despite the PSF variations introduced by the Earth's atmosphere \citep[e.g.,][]{Schmid2022}.

Detecting a disk in scattered light provides an estimate of the amount of reflected light, but often not with the accuracy required to characterize the dust in the disk. In addition, the reflected intensity depends not only on the dust scattering properties but also on the disk geometry \citep{Min2012, Ma2022}. 
Therefore, color measurements for the reflected light are useful because one can probably often assume that the scattering geometry is the same or at least very similar for different wavelengths. In this case, the color of the reflected light predominantly depends on the dust properties and not on the disk geometry. 
However, determining useful spectral gradients for the scattered light requires challenging measuring accuracies of $\Delta (Q_{\varphi}/I_{\star}) \lesssim 0.2\cdot Q_{\varphi}/I_{\star}$ or $\Delta(I_{\rm disk}/I_{\star}) \lesssim 0.2\cdot I_{\rm disk}/I_{\star}$. This can only be achieved with the analysis of high-quality data for current standards and for favorable, bright, and extended circumstellar disks. 

Up to now, successful HST color measurements for the scattered intensity from protoplanetary disks $(I_{\rm disk}/I_\star)_\lambda$ have revealed a red color for GG Tau \citep{Krist2005} and for HD~100546 \citep{Mulders2013}, while a gray color was derived for TW~Hya \citep{Debes2013}. 
With AO imaging polarimetry, estimates of the near-IR color $(Q_\varphi/I_\star)_\lambda$ were also presented but without a detailed assessment of measuring uncertainties. For seven protoplanetary disks in \cite{Fukagawa2010} and eight disks in \cite{Avenhaus2018}, the results are all, except for one disk, compatible with a gray color, or at least not strongly red or blue. The exception is the disk around IM Lup, for which a red color is reported. 
In more recent studies, predominantly red colors have been found for the polarized scattered light $(Q_\varphi/I_\star)_\lambda$ in the visible to near-IR wavelength range. This includes the measurements of HD~142527, HD~169142, and RXJ 1604 \citep{Hunziker2021, Tschudi2021, Ma2023}, as well as the estimates for HD~135344B \citep{Stolker2016, Stolker2017} and HD~34700 \citep{Monnier2019}. 

Significantly more accurate measurements can be achieved with careful data calibrations. It is shown in \citet{Tschudi2021} that variable PSF convolution and polarimetric cancellation effects are present as a result of rapidly changing AO conditions. For AO-polarimetry, this introduces variations in the $Q_\varphi/I_\star$ measurements by up to a factor of two for compact disks $(r\approx 0.2'')$ and still $\pm 20~\%$ for larger disks $(r\approx 0.5'')$.
Fortunately, this effect can be corrected using high-quality PSF calibrations and this also provides accurate estimates on the residual measuring uncertainty. Such quantitative determinations for the $(Q_\varphi/I_\star)_\lambda$ colors are available for the three disks, HD~142527, HD~169142, and RXJ 1604, mentioned above. This work now provides a larger sample of high-quality measurements for 11 disks, including three disks measured previously.

There exist dust model predictions for the expected color of the reflected light from circumstellar disks. \cite{Mulders2013} investigated dust scattering properties as a function of wavelength based on Mie theory. Small, intermediate, and large particles are expected to exhibit blue, gray, and red colors respectively. To explain the reddish color and low intensity of the reflected light from the HD~100546 disk, they proposed micron-sized porous dust aggregates, where the large particles cause the red color and produce strong forward scattering reducing the effective albedo. \cite{Tazaki2019} simulated disk colors using highly porous aggregates composed of small monomers, and compact large aggregates and they find that large aggregates composed of submicron-sized monomers can best explain gray colors. However, it seems to be far from established that the adopted models represent the dust in protoplanetary disks well, and new high-quality disk data are certainly useful to constrain better the dust parameters.  

In this study, we present accurate measurements of the polarized scattered flux $(Q_\varphi/I_\star)$ for 11 circumstellar disks observed with SPHERE at the Very Large Telescope (VLT) in the visible and near-IR wavelength, from which we derive colors or wavelength gradients $(Q_\varphi/I_\star)_\lambda$. We then try to constrain the dust properties based on the observed colors and search for systematic color trends as a function of system parameters that could be useful to infer dust evolution processes in protoplanetary disks. 
In Sect.~\ref{sect: observations} we describe the selected archival disk data and the data reduction. In Sect.~\ref{sect: cancellation} we introduce the correction method for the PSF smearing and polarization cancellation effects and then apply this correction method to the individual disk data to derive the intrinsic, disk-integrated polarized intensity in Sect.~\ref{sect:measurements}. We derive the wavelength gradients or colors for the polarized intensity in Sect.~\ref{sect:colors} and discuss the constraints provided by the observed colors on the dust properties in Sect.~\ref{sect:interpretations}. Finally, we summarize our main conclusions in Sect.~\ref{sect:conclusions}.

\begin{table*}[]
    \centering
    \caption{Stellar and disk parameters for the studied objects.}
    \label{tab: sample}  
    \resizebox{\textwidth}{!}{%
    \begin{tabular}{c c c c c c c c c c c c}
    \hline\hline
    \noalign{\smallskip}
    Host star & Type\tablefootmark{a} & $L_{\star}$\tablefootmark{a} & $M_{\star}$\tablefootmark{a} & Age\tablefootmark{a} & d\tablefootmark{b} & Incl\tablefootmark{c}  & PA\tablefootmark{c}  & $R_{\rm dust}$\tablefootmark{c}  & $F_{NIR}/F_{\star}$\tablefootmark{d} & $F_{FIR}/F_{\star}$\tablefootmark{d} & Ref. \\
            &    &  $(L_\odot)$ & $(M_{\odot})$ & (Myr) & (pc) & ($^\circ$) & ($^\circ$) & (AU) & (\%) & (\%) \\
    \hline
    \smallskip
    PDS~70    & T-Tauri K7    & 0.31 $\pm$ 0.02 & 0.8 $\pm$ 0.1    &  $7.9_{-2.9}^{+3.1}$   & 113    & 49.7      & 68.6       & 54     & $4.8\pm 0.6$  & $11.3\pm 0.7$   & 1\\
    \smallskip
    TW~Hya    & T-Tauri K6    & 0.33 $\pm$ 0.02 & 0.8 $\pm$ 0.1    & $6.3_{-1.9}^{+3.7}$    & 60     & 7         & 61/241\tablefootmark{e}     & 115    & $1.0\pm 1.0$  & $14.0\pm 0.9$   & 2, 3, 4\\
    \smallskip
    RX~J1604  & AA Tau K2     & 0.60 $\pm$ 0.03 & 1.0 $\pm$ 0.1    & $11.1_{-3.1}^{+3.3}$   & 150    & 6        & 170/350\tablefootmark{e}      & 65     & $17.5\pm 3.6$  & $26.9\pm 1.9$   & 5 \\
    \smallskip
    RX~J1615  & T-Tauri K5    & 0.90 $\pm$ 0.02 & 0.6 $\pm$ 0.1    & $1.0_{-0.2}^{+0.4}$    & 155    & 47        & 236        & 197    & $0.0\pm 0.9$   & $8.7\pm 1.0$    & 6, 7 \\
    \smallskip
    LkCa~15   & Orion Var K5  & 1.11 $\pm$ 0.04 & 1.2 $\pm$ 0.1    & $6.3_{-1.9}^{+3.7}$    & 158    & 50        & 150        & 47     & $13.4\pm 1.0$  & $9.5\pm 0.4$    & 8, 9\\
    \smallskip
    PDS~66    & T-Tauri K1    & 1.26 $\pm$ 0.04 & 1.2 $\pm$ 0.2    & $3.1_{-0.9}^{+0.8}$    & 86      & 31        & 100        & 80     & $7.3\pm 1.4$   & $6.5\pm 1.2$   & 10\\
    \smallskip

    HD~169142 & Herbig F1     & 5.6 $\pm$ 1.2   & 1.5 $\pm$ 0.2    & $12.3_{-1.2}^{+6.4}$   & 114    & 12.5      & 275 & 25   & $10.5\pm 1.0$  & $18.2\pm 3.1$   & 11 \\
    \smallskip
    HD~135344B& Herbig F8        & 6.9 $\pm$ 0.4   & 1.5 $\pm$ 0.1    & $8.9_{-1.1}^{+2.1}$    & 136    & 11        & 332       & 25      & $27.2\pm 3.1$  & $25.6\pm 1.2$   & 12\\
    \smallskip
    HD~100453 & Herbig A9     & 8.5 $\pm$ 3.5    & 1.6 $\pm$ 0.1              & $10.0_{-2.8}^{+6.3}$         & 104    & 38        & 52        & 20      & $21.7\pm 2.7$  & $19.6\pm 3.5$   & 13, 14\\
    \smallskip
    MWC~758   & Herbig A8     & 11.8 $\pm$ 0.5  & 1.8 $\pm$ 0.1    & $8.9_{-1.0}^{+2.1}$    & 160    & 21        & 155       & 70      & $27.5\pm 2.9$  & $13.1\pm 0.5$  & 15\\
    \smallskip
    HD~142527 & Herbig F6     & 20 $\pm$ 2      & 2.2 $\pm$ 0.3    & $5_{-3}^{+8}$          & 157    & 24       & 70        & 130     & $34.2\pm 3.3$  & $34.8\pm 4.1$  & 16 \\
    \hline
    \end{tabular}}
    \tablefoot{The columns give the name, stellar type, luminosity, mass, age, and distance of the host star, followed by the disk parameters
    inclination, position angle of the far side of the minor axis, the typical radius for the measured dust reflection, the near-IR excess, and the far-IR excess, and references to previous polarimetric imaging. \tablefoottext{a}{taken from \cite{Asensio-Torres2021}, except for HD142527 and HD100453 taken from their references respectively;} \tablefoottext{b}{from \cite{VanderMarel2023};} \tablefoottext{c}{taken from References;} \tablefoottext{d}{from \cite{Garufi2018};}  \tablefoottext{e}{not accurately known because of low inclination.} \\ 
    \textbf{References.} 1. \cite{Hashimoto2012}; 2. \cite{Qi2004}; 3. \cite{VanBoekel2017}; 4. \cite{Sokal2018}; 5. \cite{Pinilla2018}; 6. \cite{Pinilla2018b}; 7. \cite{DeBoer2016};  8. \cite{Thalmann2015}; 9. \cite{Thalmann2016}; 10. \cite{Wolff2016}; 11. \cite{Quanz2013}; 12. \cite{Stolker2016};  13. \cite{Benisty2017}; 14. \cite{Fairlamb2015}; 15.\cite{Benisty2015}; 16.\cite{Verhoeff2011}. 
    }
\end{table*}

\section{Observations and data reduction}
\label{sect: observations}
\subsection{Target selection}
The study conducted in the paper is based on SPHERE data from the European Southern Observatory (ESO) archive. We specifically selected protoplanetary disks that were observed with both the visible Zurich imaging polarimeter (ZIMPOL) and the near-IR dual-band imager and spectrograph (IRDIS) focal plane instruments of SPHERE \citep{Beuzit2019} to facilitate a multiwavelength analysis. As determining the polarized flux of protoplanetary disks is challenging, we only considered high-quality data from bright, extended transition disks or pre-transition disks \citep{Espaillat2014, VanderMarel2023}. The observations were required to include unsaturated flux observations of the central star, which serve as a photometric reference for our measurements and as a PSF calibrator to assess the AO performance.

\begin{figure*}
    \centering
    \includegraphics[width=0.9\textwidth]{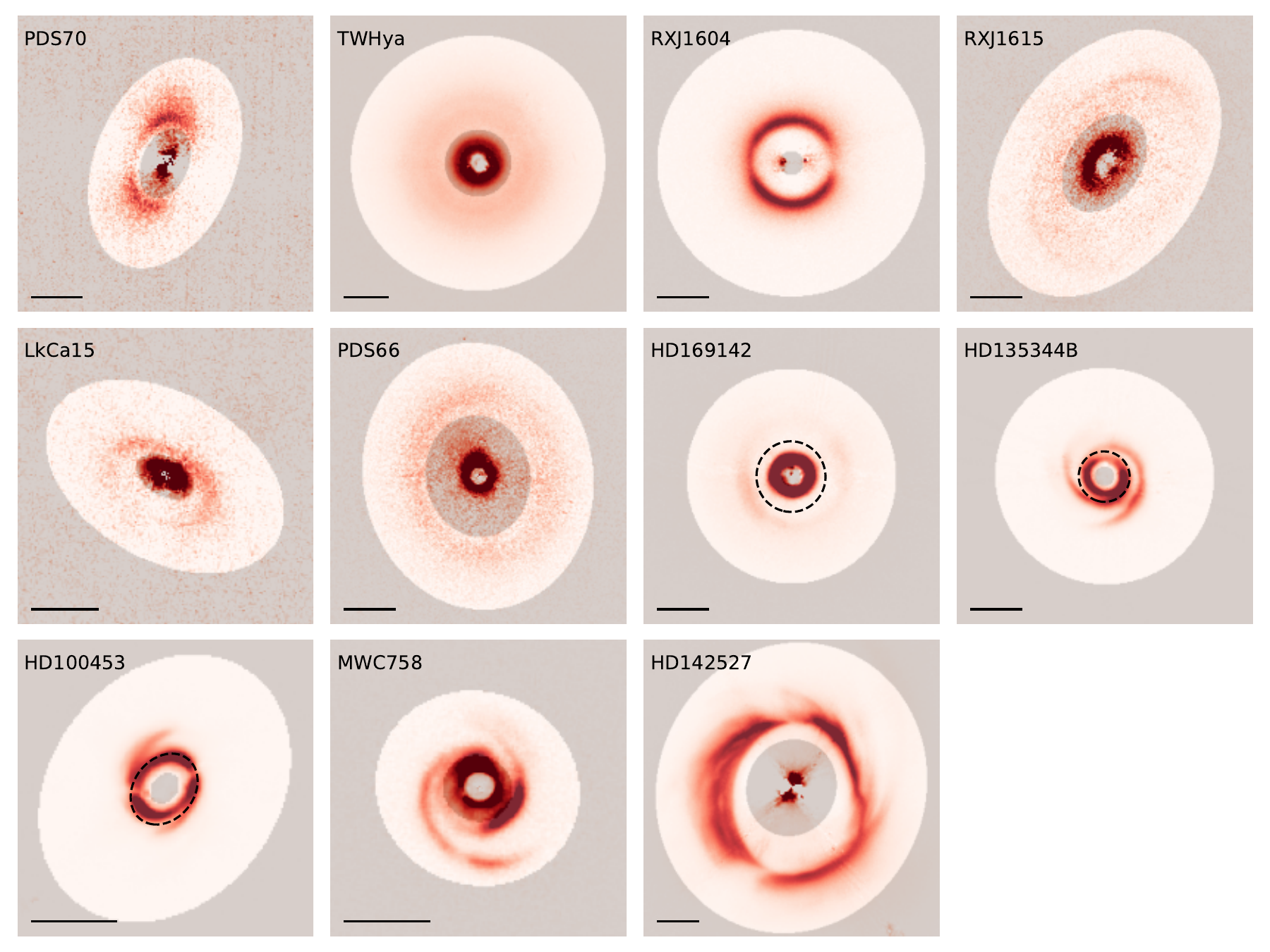}
    \caption{Observed polarized intensity images $Q_\varphi(x,y)$ of the studied disks. $J$ band images are given, except for TW~Hya which is shown in the H band and MWC~758 in the $Y$ band. The lines in the left-bottom corner represent 0.5\arcsec. The white elliptical annulus highlights the disk integration region and the dashed ellipses for HD~169142, HD~135344B, and HD~100453 separate the inner rings from the outer disk structures (Sect.~\ref{Sect:in-out}).
    }
    \label{fig:gallery}
\end{figure*}

Polarimetric images of the selected 11 targets are shown in Fig.~\ref{fig:gallery} with the flux integration region outlined. All targets show a bright disk rim or a strong ring structure with a well-resolved and clearly defined inner boundary. Some of the disks include fainter outer structures, such as spirals, which are also integrated in the disk flux. Our sample consists of six T-Tauri stars and five Herbig stars, with disk inclinations ranging from $i=6^\circ$ to $50^\circ$. We derive $Q_\varphi/I_\star$ values using the same "generalized" PSF correction and disk measuring procedure for all our targets. 
Our sample includes three targets RX~J1604, HD~169142, and HD~142527, which have previous quantitative $Q_\varphi/I_\star$ measurements \citep{Ma2023, Tschudi2021, Hunziker2021}. These three targets were remeasured using our "generalized" correction procedure for consistency and for a quality assessment of our analysis. 

We provide a summary of the important system properties of our targets in Table.~\ref{tab: sample}, which includes references to previous descriptions of the polarized scattered light signal of the disk. 
The archival datasets used in our study are listed Table.~\ref{tab:obs-info} including information on filters, instrument settings, integration times, and observing conditions. The near-IR images taken with IRDIS have a much larger pixel scale of $12.25~{\rm mas}\times 12.25~{\rm mas}$ and field of view (FoV) of $10'' \times 12.5''$ than the visible data from ZIMPOL with pixels of $3.6~{\rm mas}\times 3.6~{\rm mas}$ and FoV of $3.6'' \times 3.6''$.
Therefore, the disk integration range is for all disks smaller than the FoV of ZIMPOL, while parameters expressed in pixels always refer to the size of an IRDIS pixel.

\subsection{Data reduction}
The ZIMPOL data were subjected to basic data reduction steps using the sz-pipeline developed at ETH Zurich as described in \citet{Schmid2018}, and the IRDIS data were processed using the IRDAP pipeline \citep{VanHolstein2020}. The basic reduction steps include data extraction, bias subtraction (ZIMPOL) or dark subtraction(IRDIS), flat fielding, polarimetric combination, calibration of the polarimetric efficiency, and corrections for the polarimetric beam-shift effect (ZIMPOL only). 
The central stars in most systems show no or only a small intrinsic or interstellar polarization ($p\leq 0.2~\%$) and thus a polarimetric normalization is applied. A significant intrinsic polarization, probably due to an unresolved and inclined inner disk is recognized for the central star of a few objects PDS~70, LkCa~15, RXJ~1615 \citep{Keppler18, Thalmann2015, DeBoer2016}, and this is then corrected in an additional calibration step. \\
Because the scattered light from the disk is linearly polarized in the azimuthal direction, we convert the reduced Stocks polarizations $Q(x,y)$ and $U(x,y)$ to the azimuthal polarization parameters $Q_{\varphi}(x,y)$ and $U_{\varphi}(x,y)$ with:
\begin{align}
    Q_{\varphi}(x,y) & = -Q(x,y)\cos(2\varphi) - U(x,y)\sin(2\varphi) \\
    U_{\varphi}(x,y) & = Q(x,y)\sin(2\varphi) - U(x,y)\cos(2\varphi)
\end{align}
where $\varphi$ is the position angle of point $(x,y)$ with respect to the central star from north over east. The polarization in azimuthal direction $Q_{\varphi}(x,y)$ represents the observed polarized disk intensity image. \\
The non-coronagraphic and non-saturated calibration frames of the central star were used for photometric calibrations and the PSF assessments. No calibration frames were used for polarimetric disk observations taken in non-coronagraphic mode and without saturation of the central star because the intensity signal of the star can be used for the calibration. More details of the instrument configurations and the data reduction procedures are provided in the Appendix (Sect.~\ref{tab:obsdetails}), and non-standard procedures applied to special targets are outlined in Sect.~\ref{sect: special}.


\section{Correction for the PSF convolution}
\label{sect: cancellation}
The correction for the PSF convolution effects is a crucial step in determining the polarized flux of a circumstellar disk. For the SPHERE AO system, the PSF for the visible observation with ZIMPOL typically has a narrow, coherent light core containing 10-50~\% of all light, with a full width at half maximum of about 0.02 - 0.03~arcsec, while for IRDIS, the core contains 40-80~\% of the light with a width of 0.03 - 0.05~arcsec. 
The rest of the flux is distributed in an extended residual seeing halo with a width of about 0.5 - 1 arcsec. Convolution with such a PSF causes a smearing of the intrinsic disk signal and polarimetric cancellation effects between regions with positive and negative $Q$ signals, and positive and negative $U$ signals \citep{Schmid2006}. Each instrument has its individual PSF, which varies systematically with wavelength, and for ground-based AO observations, the PSF varies considerably on short time scales with the atmospheric observing conditions \citep{Tschudi2021}. For our disk data, the $Q_\varphi$ flux degradation effects are 20~\% or larger and sometimes even reach a factor of two. Thus, accurate determination of the polarized scattered light and its wavelength dependence requires an assessment and correction for these effects. 

To properly account for the PSF convolution effects, a good method is to define a parametric disk model and search for the best disk parameter by comparing many PSF-convolved disk models with the observations. This approach is quite straightforward for ring-shaped protoplanetary disks like HD~169142 \citep{Tschudi2021}, RX~J1604 \citep{Ma2023} or for debris disk rings HR~4796A \citep{Chen2020}, HD~141569 \citep{Singh2021},  HD~129590 \citep{Olofsson2023}. However, more complicated disks require many parameters for a model description of their geometry. A good example, which we also use in this work as a test case (Fig.~\ref{fig:hd100453}), is the disk around HD~100453 consisting of a bright ring, two spiral arms, and two shadows from an inner inclined disk, which was modeled in \citet{Benisty2017} with more than 20 disk geometry parameters. 
This approach seems impractical for a heterogenous disk sample, as presented in this work. Therefore, we investigate and use a new approximate correction procedure for the PSF convolution effects in our measurements.

\begin{figure*}
    \centering
    \includegraphics[width=0.8\textwidth]{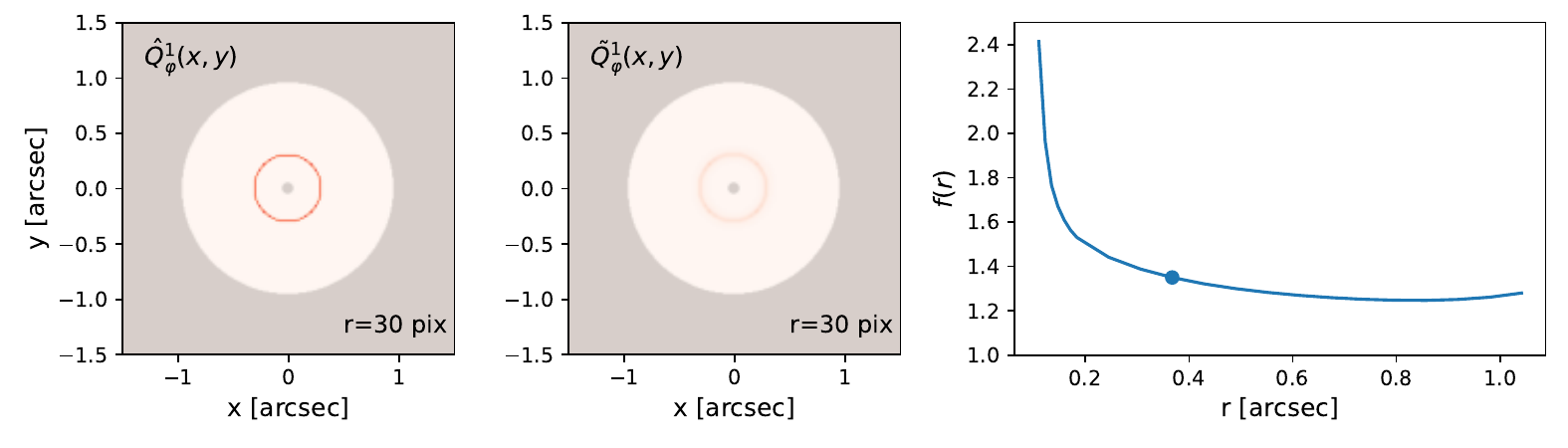}
    \caption{Illustration of correction procedure with 1-pix wide rings. Left: 1-pix wide ring $\hat{Q}^{1}_{\varphi}(x, y)$ with radius $r=30$ pix or $r = 0.37\arcsec$ in IRDIS pixel scale. Middle: Convolved image $\tilde{Q}^{1}_{\varphi}(x, y)$ with radius $r=30$ pix, obtained after the convolution process with IRDIS-$J$ band PSF from 16-06-22. Right: Correction function $f(r)$. The blue dot indicates the local correction factor at 30 pix calculated from the left images. }
    \label{fig: corr_curve}
\end{figure*}

\subsection{Concept for the correction procedure}
We describe first our procedure for the correction of the PSF convolution effects for an axis-symmetric disk seen pole-on using radial correction curves $f(r)$. For inclined disks one needs then to extend this approach to 2-dimensional correction maps $f(x,y)$. 
Convolution is a linear operation and therefore the convolution of an azimuthally symmetric disk can be obtained as the sum of convolutions of narrow azimuthal rings. 
We start with the convolution of a model ring $\hat{Q}^1_\varphi(r,x,y)$\footnote[1]{For the polarization parameters $Q, U, Q_{\varphi}$ and $U_{\varphi}$, we distinguish between the plain symbols for the observed quantities, symbols with a hat (e.g., $\hat{Q}_{\varphi}$) for the PSF smearing corrected intrinsic quantities, and symbols with a tilde (e.g., $\tilde{Q}_{\varphi}$) for the PSF convolved model parameters. The superscript 1 refers to the one-pixel ring used for the convolution modeling. The symbol with a bar is only used for test disk models. } with radius $r$, a width of one pixel with a "normalized" flux of one, and a total flux equal to the number of nonzero pixels tracing the ring ($\hat{Q}^1_\varphi(r) = \sum_{x,y} \hat{Q}^1_\varphi(r,x,y) \approx {\rm int}(2\pi r)$). We assume, that the intrinsic disk polarization has a perfect azimuthal polarization ($\hat{U}_\varphi=0$). 

The $\hat{Q}^1_\varphi$-ring is converted into intrinsic Stokes signal rings $\hat{Q}^1(r,x,y)$ and $\hat{U}^1(r,x,y)$ with an azimuthal dependence of $\hat{Q}^1(\theta_{xy}) =-\cos (2\theta_{xy})$ and $\hat{U}^1(\theta_{xy}) =\sin (2\theta_{xy})$ with positive and negative quadrants. These frames are then convolved with the PSF of the observations into $\Tilde{Q}^1(r,x,y)$ and $\Tilde{U}^1(r,x,y)$. The resulting signal is now radially spread into a broader ring and a halo with a quadrant pattern and the integrated Stokes fluxes are reduced $\Tilde{Q}^1(r)<\hat{Q}^1(r)$ and $\Tilde{U}^1(r)<\hat{U}^1(r)$, because of cancellation between positive and negative signals, and because some of the signals are distributed outside of the integration region. 
The convolved Stokes flux images are then converted to convolved azimuthal polarization images $\Tilde{Q}^1_\varphi(r,x,y)$ and $\Tilde{U}^1_\varphi(r,x,y)$. Because of the convolution, the $\Tilde{U}^1_\varphi(r,x,y)$-signal now differs from zero, and therefore the integrated disk signal $\Tilde{Q}^1_\varphi(r)$ is further reduced. 

The cancellation effect at radius $r$ for the azimuthal polarization can be approximated by the ratio between the integrated intrinsic flux of the pixel ring $\hat{Q}^1_\varphi(r)$ and the integrated convolved flux for this pixel ring $\Tilde{Q}^1_\varphi(r)$ in the chosen integration region from $r_1$ to $r_2$.
\begin{equation}
    f(r) = \hat{Q}^1_\varphi(r)/\Tilde{Q}^1_\varphi(r)\,.
\end{equation}
This radial function defines an axisymmetric correction map $f(x,y)$ which can be applied to the observed azimuthal polarization image $Q_\varphi(x,y)$ to recover the intrinsic disk flux.
\begin{equation}
    \hat{Q}_\varphi = \sum_{x,y} \hat{Q}_\varphi(x,y) \approx \sum_{x,y} f(x,y) Q_\varphi(x,y)\,.
\end{equation}
This axisymmetric map does not depend on the disk brightness distribution. It calculates how much disk flux is typically lost at a given radius $r$ on a relative scale. 
The focus of this procedure is to recover the integrated, intrinsic disk polarization signal. It is sufficiently accurate for disks without strong azimuthal structures and without very strong radial structures.

An illustration of the construction and the typical shape of a radial correction curve $f(r)$ is illustrated in Fig.~\ref{fig: corr_curve} using the IRDIS-$J$ band PSF from Jun 22, 2016 in a $244\times 244$ pixels grid to simulate a $3\arcsec \times 3 \arcsec$ field of view in IRDIS. The inner and outer integration borders are $r_{\rm in}=8$~pix ($0.10"$) and $r_{\rm out}=95$~pix ($1.16"$), which correspond approximately to the integration region for the low inclination (close to pole-on) disk HD~135344B. 
The correction curve $f(r)$ has high values at the inner border where the smearing by the PSF-peak already introduces strong polarimetric cancellation of a disk signal in addition to a significant flux loss into the central region $r<r_{\rm in}$, which is not considered in the flux integration. Further out, polarization cancellation results mainly from the flux smearing by the extended PSF halo, and therefore the curves are rather flat at a level of 1.3 or about 30~\% loss in polarization flux. At the outer integration border the curve rises again slightly because more smeared flux is lost to regions outside the integration range $r>r_{\rm out}$.

\begin{figure}
    \centering
    \begin{subfigure}[]{0.49\textwidth}
        \includegraphics[width=\textwidth]{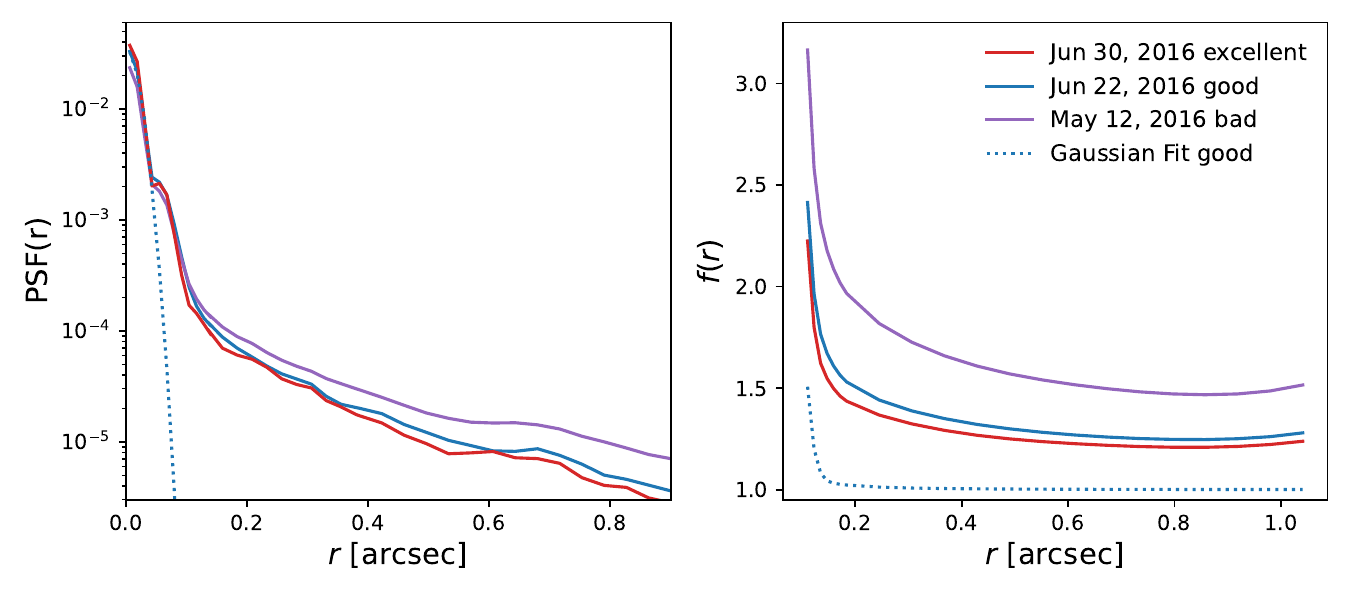}
    \end{subfigure}
    \begin{subfigure}[]{0.49\textwidth}
        \includegraphics[width=\textwidth]{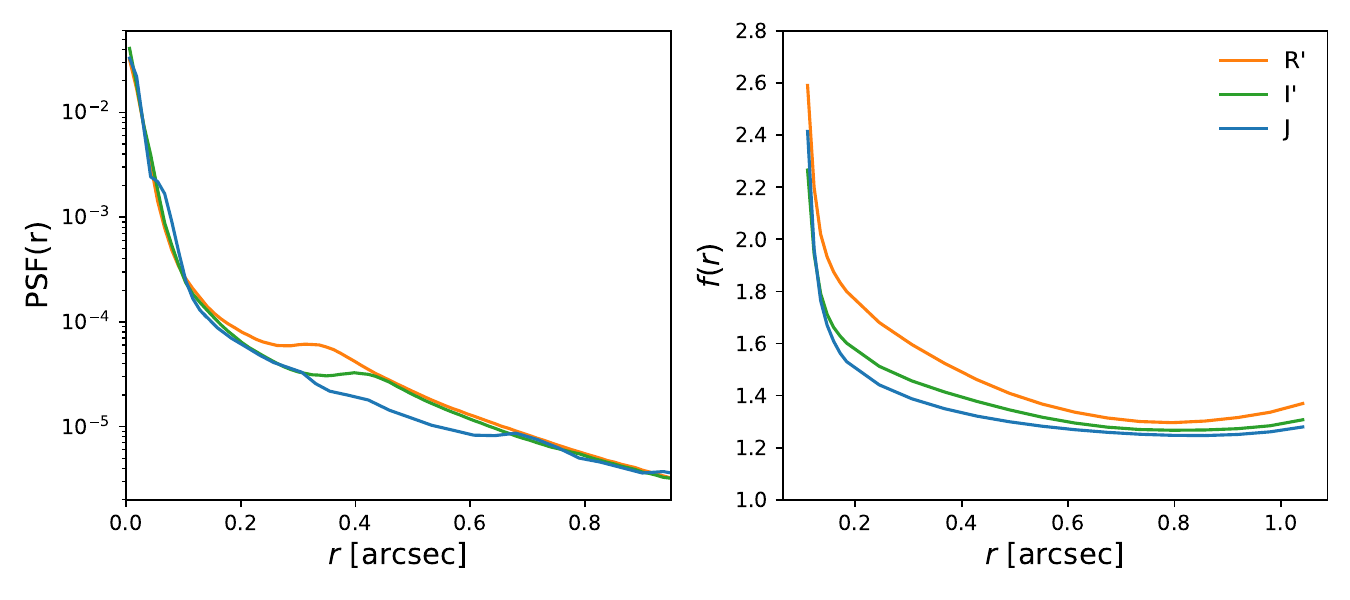}
    \end{subfigure}
    \caption{Typical PSFs and correction curves for different observing conditions and wavelengths. Upper: Typical PSFs and correction curves $f(r)$ for the IRDIS J-band for "excellent", "good", and "bad" conditions and a Gaussian fit to the PSF peak for "good" condition. Lower: PSFs and $f(r)$ at different wavelengths for 
    good quality conditions.}
    \label{fig:corr-psf-dep}
\end{figure}

\begin{figure*}
    \centering
    \includegraphics[width=0.8\textwidth]{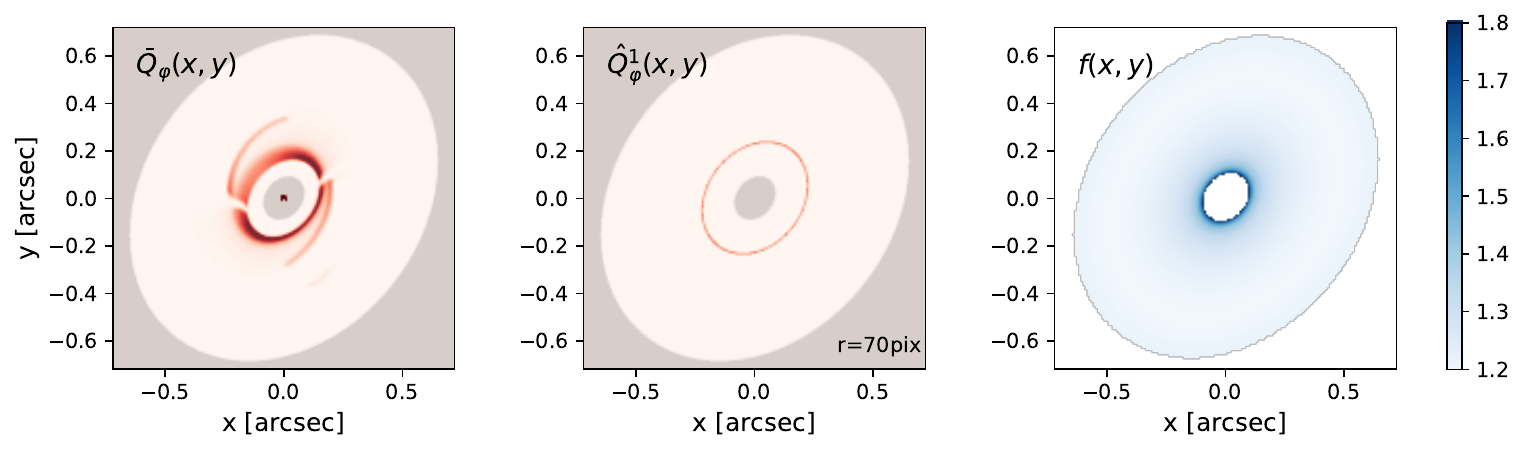}
    \caption{Illustration of the correction method for HD~100453. Left: Test model $\bar{Q}_{\varphi}(x,y)$ for HD~100453 from \cite{Benisty2017}. 
    Middle: Inclined, 1-pixel wide polarized intensity ring $\hat{Q}^1_{\varphi}(r,i,\phi,x,y)$. 
    Right: Correction map $f(x,y)$ defined by $i$, $\phi$, $r_{\rm in}=0.10\arcsec$ and $r_{\rm out} = 0.74\arcsec $, simulated in ZIMPOL pixel scale.
    }
    \label{fig:hd100453}
\end{figure*}

Knowing well the PSF of the observations is most important to account for the smearing effects and to deriving the correction maps $f(x,y)$, which quantify accurately the flux loss. Figure~\ref{fig:corr-psf-dep} illustrates the PSF dependence of the radial correction curves $f(r)$ on the PSF for the pole-on disk model. The graphs show the normalized PSFs and the $f(r)$-curves obtained for $J$ band representing excellent, good and bad atmospheric conditions, and PSFs and $f(r)$ for wavelengths at 626~nm (ZIMPOL $R'$), 790~nm (ZIMPOL $I'$) and 1245nm (IRDIS $J$) for good atmospheric conditions. The PSFs in the first column are normalized to the total counts in the 3'' round aperture and the corresponding observing conditions and $f(r)$-characteristics are given in Table~\ref{tab:psf_quality}.
Moreover, we fit a 2D Gaussian to the peak of the J-band good condition and calculate $f(r)$ shown as the dotted curve in Fig.~\ref{fig:corr-psf-dep}. The Gaussian approximation ignores the halo and it results in an $f(r)$ correction curve which converges to 1 quickly, and therefore significantly underestimates the correction for the intrinsic flux. 
The examples in Fig.~\ref{fig:corr-psf-dep} illustrate that the correction factors $f(r)$ differ significantly from case to case and an accurate derivation of the PSF for each observation is essential.

\begin{table*}[]
    \centering
    \caption{PSF and correction curve $f(r)$ characteristics for a few representative cases for SPHERE}
    \begin{tabular}{c c c c c c c c c c}
    \hline
    \hline
    Date   & \multicolumn{3}{c}{Atmosphere}             & \multicolumn{2}{c}{Instrument mode}  & \multicolumn{3}{c}{$f(r)$ charateristics} \\
    Category & seeing($\arcsec$) & $\tau_0(ms)$ & airmass &   $n_{DIT} \times t_{DIT}$ & Filter    & $r=0.15\arcsec$ & $0.30\arcsec$ & $0.60\arcsec$  \\
    \hline
    "excellent"\\
    Jun 30, 2016  &    0.41    &  5.5           & 1.03    & 4 $\times$ 0.8375 s & J & 1.53 & 1.33 & 1.23\\
    "good"\\
    Jun 22, 2016  &    0.73    &  5.1           & 1.16    & 4 $\times$ 0.8375 s & J & 1.66 & 1.39 & 1.27\\
    Apr 01, 2015  &   0.74     & 3.0            & 1.10    & 4 $\times$ 10s      & I' & 1.70 & 1.46 & 1.30\\
              &   0.74     & 3.0            & 1.10    & 4 $\times$ 10s      & R' & 1.92 & 1.60 & 1.34\\
    "bad"\\
    May 12, 2016  &    1.96    &  1.1           & 1.13    & 10 $\times$ 2 s     & J & 2.15 & 1.74 & 1.52\\
    
    \hline
    \end{tabular}
\tablefoot{The data were selected from the PSF calibrations obtained for HD~135344B 
    (see Table~\ref{tab:obs-info}). The contribution of the
    disk intensity to the stellar PSF is estimated to be about 2~\% for the R' and I' bands, 
    and about 4~\% for the $J$ band assuming $I_{\rm disk}\approx 4\cdot 
    \hat{Q}_\varphi$.}
    \label{tab:psf_quality}
\end{table*}

\subsection{Correction maps for inclined disks}
\label{sect:corr-map}
Disks are in most cases significantly inclined and therefore we extend the correction procedure to two-dimensional correction map based on one pixel wide ellipses $\hat{Q}^1_\varphi(r,i,\phi, x, y)$ representing inclined rings.
The pixel coordinates with a value one are defined by
\begin{align}
x = &  +r \cos(\theta) \cos(\phi) +  r \cos(i) \sin(\theta) \sin(\phi) \,\\
y = &  -r \cos(\theta) \sin(\phi) + r \cos(i) \sin(\theta) \cos(\phi) \,
\end{align}
where $r$ is the semi-major axis of the ellipse, $i$ is the inclination, and $\phi$ is the orientation of the semi-minor axis measured from N to E (counterclockwise). All other pixels are set to zero. The same convolution as for pole-on rings is then applied to this ellipse. The integration region is defined by inclined rings with the same $i$ for the inner and outer borders with projected semi-major axis $r_1$ and $r_2$. The ratio of the total fluxes between intrinsic and convolved polarization yields the correction value for all pixels of the considered ellipse with semi-major axis $r$.
\begin{equation}
f(r,i,\phi) = \hat{Q}^1_\varphi(r,i,\phi)/\tilde{Q}^1_\varphi(r,i,\phi)\,.
\end{equation}
This procedure is then repeated for all elliptical rings with $r_1<r<r_2$ for the construction of an elliptical correction map $f(x,y)$ 
equivalent to the pole-on case discussed above. 
As an example, Fig.~\ref{fig:hd100453} shows the inclined disk HD~100453 in the $I'$ band, for which we can use the detailed model described in \citet{Benisty2017} for the intrinsic disk signal. The figure also shows an elliptical disk ring $\hat{Q}_\varphi^1(r, i, \phi)$ with flux one and a width of one pixel, and the final correction map $f(x,y)$ for the corresponding $I'$ band PSF. 
Correcting the convolved $\tilde{Q}_\varphi(x,y)$ polarized flux model image of HD~100453 shown in Fig.~\ref{fig:hd100453} with this $f(x,y)$-map yields an intrinsic disk signal which only differs by 0.3~\% from the value derived for the test model $\bar{Q}_\varphi$. 
The derivation is based on noise-free models. Adding pixel-to-pixel photon noise and read-out noise produces 5\% uncertainty in the integrated $\hat{Q}_{\varphi}$ for the faintest disks, and much less for bright disks.
Additional tests for this correction are described in the Appendix and we find that the typical systematic uncertainty introduced by this procedure is at the level of 1-2\%.

\subsection{PSF selection}
For each disk observation, the selection of the PSF is very important for the derivation of the correction map and the determination of the intrinsic polarized flux $\hat{Q}_\varphi$.
We always use an intensity PSF of the central star for the correction of the $Q_\varphi$ convolution which also includes the intensity of the disk. This contamination of the stellar PSF is small and can be taken into account for the final $\hat{Q}_\varphi/I_\star$ value as will be discussed in the next section. Using the central star as a reference has the significant advantage, that PSF and imaging polarimetry of the disk are taken shortly after each other or even simultaneously so that temporal variations of the atmospheric turbulence and corresponding changes in the AO performance and the PSF profile are minimized. In addition, using the central star as a PSF reference is also an efficient strategy in terms of observing overheads. 

The PSF selection and determination procedure for the individual data depends on the instrument configuration and the applied observing strategy. We distinguish between three quality classes for the PSF calibration which are described in the following.

\paragraph{A. Simultaneous PSF.}
Non-coronagraphic observations without saturation of the intensity signal of the central star provide simultaneously the Stokes parameters $Q=I_0-I_{90}$ and $U=I_{45}-I_{135}$ of the polarized radiation from the disk, as well as the corresponding intensity images $I_Q=I_0+I_{90}$ and $I_U=I_{45}+I_{135}$ for each polarimetric cycle. We use the mean intensity $I_{\star} = (I_Q+I_U)/2$ for the PSF calibration. Changes in PSF introduced by atmospheric variations can be considered cycle by cycle with the corresponding correction map $f(x,y)$. 
Such disk measurements are available for many ZIMPOL datasets and for the IRDIS $J$ band data of the targets LkCa~15 and PDS~70, which are faint enough to be observed without coronagraph and saturation. 

\paragraph{B. Non-simultaneous PSF.}
This category includes datasets where the central star is saturated or hidden behind the coronagraphic mask to obtain deep polarimetric images of the circumstellar disk. For the flux and PSF calibration, separate exposures of the central star are taken, without coronagraph or with the star offset from the mask, using short integration times and/or neutral density (ND) filters to avoid detector saturation. These data are not taken strictly simultaneously, but immediately before or/and after the disk observations so that small changes in the PSF may have occurred. 

We consider two sub-cases of data calibration: class B1 for datasets with only one useful flux or PSF dataset taken either before or after all the polarimetric cycles. This PSF is then used for the correction map, which is applied to all data of one target taken in a given filter. 
In this case, we have no or only very limited information about short-term flux and PSF variability and we adopt uncertainties in the PSF profile at the level of $\sigma(I_{\star}) = 5\%$. 
Class B2 are datasets for which flux frames are taken more than once, typically before and after the polarimetric cycles. 
We use in this case the average of the calibration data and generate a single calibration map as in the B1 case, which is applied to all data. However, having more than one calibration set provides a useful check for the stellar flux and PSF variability. There remains a lack of statistics but if the stellar flux does not change between these frames then we adopt $\sigma(I_{\star}) = 3\%$. Otherwise, we discuss the uncertainties in Sect. \ref{sect: special} under special cases.  

\paragraph{C. Multi-epoch PSFs}
This class consists of multi-epoch B-type observations for a given target and filter. Such datasets were obtained for the monitoring of variable shadows in RX~J1604 \citep{Pinilla2018} and HD~135344B \citep{Stolker2017}. The datasets are split into several observing runs including PSF calibrations for each epoch. The H-band data of HD~142527 were all taken in the same night but PSFs were taken several times. 
We split these data and treat the parts like separate runs, each with its own PSF calibration.
We apply the B-type correction process to each epoch and derive the mean $\hat{Q}_{\varphi}/I_\star$ and standard deviations for all high-quality runs similar to the study of \citet{Ma2023} for RX~J1604. 
This yields a higher precision when compared to B-type single-run data. 

\begin{table*}[]
    \centering
    \caption{Measurement results for the selected disks. }
    \begin{tabular}{c|c c c c c c c}
    \hline
    \hline
    Host star  
        & Filt 
            & $I_{\star}$ [$10^6$ct/s]  
                & $Q_{\varphi}$ [$10^4$ct/s]  
                    & $\hat{Q}_{\varphi}$ [$10^4$ct/s]         
                        & $Q_{\varphi}/I_{\star}$ [\%] 
                            & $\hat{Q}_{\varphi}/I_{\star}$ [\%] 
                                & Process\\
    \hline
    PDS~70     & VBB  & 1.30 $\pm$ 0.01 & 0.20 $\pm$ 0.02 & 0.46 $\pm$ 0.01
                      & 0.15 $\pm$ 0.02 & 0.36 $\pm$ 0.01 & A\\
    $[0.36,1.10]$& J    & 1.75 $\pm$ 0.01 & 0.32 $\pm$ 0.02 & 0.57 $\pm$ 0.05
                      & 0.19 $\pm$ 0.01 & 0.32 $\pm$ 0.03 & A\\
               & H    & 1.16 $\pm$ 0.06 & 0.18 $\pm$ 0.01 & 0.28 $\pm$ 0.02 
                      & 0.15 $\pm$ 0.02 & 0.24 $\pm$ 0.02 & B1\\
               & Ks   & 0.72 $\pm$ 0.02 & 0.22 $\pm$ 0.02 & 0.29 $\pm$ 0.02
                      & 0.30 $\pm$ 0.03 & 0.40 $\pm$ 0.04 & B2 \\[0.1cm]
    TW Hydrea  & R'     & 2.14 $\pm$ 0.02 & 0.90 $\pm$ 0.04 & 1.37 $\pm$ 0.04
                        & 0.42 $\pm$ 0.02 & 0.64 $\pm$ 0.02 & A\\
    $[0.39,1.50]$& I'     & 1.95 $\pm$ 0.02 & 1.03 $\pm$ 0.02 & 1.47 $\pm$ 0.02
                        & 0.53 $\pm$ 0.01 & 0.75 $\pm$ 0.01 & A\\
               & H    & 4.1 $\pm$ 0.2 & 2.45 $\pm$ 0.04 & 2.83 $\pm$ 0.05 
                      & 0.60 $\pm$ 0.04 & 0.69 $\pm$ 0.05 & B2\\[0.1cm]
    RX~J1604   & R\_d & 0.559 $\pm$ 0.005 & 0.25 $\pm $ 0.02 & 0.500 $\pm$ 0.008 & 0.44 $\pm$ 0.04 & 0.89 $\pm$ 0.02 & A \\
    $[0.17,1.47]$& J    & 0.95 $\pm$ 0.02   & 1.10 $\pm$ 0.04 & 1.44 $\pm$ 0.04 & 1.16 $\pm$ 0.07 & 1.52 $\pm$ 0.08 & C\\[0.1cm]
    RX~J1615   & R\_d & 0.54 $\pm$ 0.01 & 0.048 $\pm$ 0.002 & 0.078 $\pm$ 0.002
                      & 0.089 $\pm$ 0.005 & 0.145 $\pm$ 0.006 & A\\
    $[0.54, 1.47]$ & J    & 0.93 $\pm$ 0.03 & 0.11 $\pm$ 0.01 & 0.17 $\pm$ 0.01
                      & 0.11 $\pm$ 0.01 & 0.18 $\pm$ 0.02 & B2\\
               & H    & 1.22 $\pm$ 0.06 & 0.187 $\pm$ 0.006 & 0.32 $\pm$ 0.01
                      & 0.15 $\pm$ 0.01 & 0.27 $\pm$ 0.02 & B1\\[0.1cm]
    LkCa~15    & VBB    & 1.28 $\pm$ 0.12 & 0.07 $\pm$ 0.01 & 0.14 $\pm$ 0.01
                        & 0.065 $\pm$ 0.006 & 0.122 $\pm$ 0.008 & A\\
    $[0.23, 0.98]$ & J  & 1.534 $\pm$ 0.004 & 0.39 $\pm$ 0.06 & 0.57 $\pm$ 0.08
                        & 0.25 $\pm$ 0.04 & 0.37 $\pm$ 0.05 & A \\
               & Ks     & 0.97 $\pm$ 0.03 & 0.30 $\pm$ 0.02 & 0.48 $\pm$ 0.03
                        & 0.30 $\pm$ 0.03 & 0.50 $\pm$ 0.03 & B2 \\[0.1cm]
    PDS~66     & N\_R & 1.34 $\pm$ 0.07 & 0.120 $\pm$ 0.007 & 0.20 $\pm$ 0.01
                      & 0.089 $\pm$ 0.010 & 0.15 $\pm$ 0.02 & B1\\
    $[0.61,1.35]$ & J    & 2.6 $\pm$ 0.1 & 0.26 $\pm$ 0.01 & 0.35 $\pm$ 0.02
                      & 0.10 $\pm$ 0.01 & 0.14 $\pm$ 0.01 & B1\\
               & H    & 4.3 $\pm$ 0.1 & 0.35 $\pm$ 0.02 & 0.40 $\pm$ 0.02
                      & 0.082 $\pm$ 0.006 & 0.093 $\pm$ 0.007 & B2\\[0.1cm]
    HD~169142  & R'    & 2.28 $\pm$ 0.02   & 0.47 $\pm$ 0.04 & 1.02 $\pm$ 0.03 & 0.20 $\pm$ 0.02 & 0.45 $\pm$ 0.02 & A\\
    $[0.10, 1.08]$& I'    & 1.42 $\pm$ 0.02   & 0.43 $\pm$ 0.04 & 0.82 $\pm$ 0.02 & 0.30 $\pm$ 0.03 & 0.57 $\pm$ 0.02 & A\\
               & J    & 5.2 $\pm$ 0.3     & 2.92 $\pm$ 0.07 & 4.14 $\pm$ 0.09 & 0.57 $\pm$ 0.04 & 0.80 $\pm$ 0.06 & B1 \\[0.1cm]
    
    HD~135344B & R'     & 10.5 $\pm$ 0.5 & 2.57 $\pm$ 0.06 & 4.26 $\pm$ 0.10
                        & 0.24 $\pm$ 0.02 & 0.40 $\pm$ 0.03 & B1\\
    $[0.10,1.15]$& I'     & 7.3 $\pm$ 0.4  & 2.65 $\pm$ 0.10 &  4.02 $\pm$ 0.15 
                        & 0.36 $\pm$ 0.03 & 0.55 $\pm$ 0.05 & B1\\
               & J      & 6.1 $\pm$ 0.3 & 4.03 $\pm$ 0.61 & 5.99 $\pm$ 0.67
                        & 0.67 $\pm$ 0.13 & 0.99 $\pm$ 0.16 & C\\[0.1cm]
    HD~100453  & R'     & 24.7 $\pm$ 2.5 & 4.37 $\pm$ 0.58 & 8.47 $\pm$ 1.13 
                        & 0.18 $\pm$ 0.04 & 0.34 $\pm$ 0.08 & B1\\
    $[0.10,0.86]$& I'     & 12.8 $\pm$ 1.3 & 3.97 $\pm$ 0.48 & 5.97 $\pm$ 0.72
                        & 0.31 $\pm$ 0.07 & 0.46 $\pm$ 0.10 & B1\\
               & J      & 7.6 $\pm$ 0.8 & 4.25 $\pm$ 0.13 & 7.17 $\pm$ 0.23
                        & 0.56 $\pm$ 0.08 & 0.94 $\pm$ 0.13 & B2\\[0.1cm]
    MWC~758    & VBB  & 3.01 $\pm$ 0.02 & 0.25 $\pm$ 0.01 & 0.52 $\pm$ 0.01
                      & 0.084 $\pm$ 0.005  &  0.174 $\pm$ 0.004 & A \\
    $[0.21, 0.61]$ & Y    & 3.9 $\pm$ 0.2 & 0.74 $\pm$0.02 & 1.30 $\pm$ 0.03
                      & 0.19 $\pm$ 0.01 & 0.33 $\pm$ 0.02 & B1\\[0.1cm]
    HD~142527  & VBB  & 4.06 $\pm$ 0.03   & 2.6 $\pm$ 0.1   & 3.3 $\pm$ 0.1 & 0.32 $\pm$ 0.02 & 0.40 $\pm$ 0.02 & A \\
    $[0.61,1.84]$& H    & 21.3 $\pm$ 0.4    & 18.1 $\pm$ 0.2  & 21.3 $\pm$ 0.2 & 0.85 $\pm$ 0.03 & 1.00 $\pm$ 0.03 & C\\
    \hline
    \end{tabular}
    \label{tab:integrate-flux}
    \tablefoot{The columns are the measurement results for the integrated intensity $I_\star$, the polarized intensity $Q_\varphi$, the convolution-corrected polarized intensity $\hat{Q}_\varphi$, the polarized disk reflectivity $Q_\varphi/I_\star$, and the corresponding, convolution corrected value $\hat{Q}_\varphi/I_\star$, and the used PSF calibration process. The considered disk integration region is given below the target name by $[r_1,r_2]$ (in arcsecs).}
\end{table*}

\section{Measurements of the polarized intensity $Q_\varphi/I_\star$}
\label{sect:measurements}
A total of 31 intensity measurements in various wavelength bands for 11 disks are presented in Table~\ref{tab:integrate-flux}. 
The total target intensity $I_\star$ and the azimuthal polarization $Q_\varphi$ are given on counts per second scale specific for each filter and instrument mode. The observed total target intensity $I_\star$ is scaled to the $Q_\varphi$-measurements for cases where the $I_\star$ is taken in a different instrument mode, like additional ND-filter or different detector gain settings (for ZIMPOL). The aperture for the $I_\star$ measurements has for all observations a diameter of $3''$ and includes therefore the intensity of the disk. 
A background level is subtracted based on the mean counts for an annulus located just outside the aperture. The uncertainty for $I_\star$ count is in most cases $\leq 5$~\% based on the measured deviations between different measurements indicating that no significant flux variations are recognized. The exceptions are the values for HD~100453 and LkCa~15 which are discussed separately.  

The integration regions for the disk polarization $Q_\varphi$ are defined by ellipses with major axes $r_1$ and $r_2$ for the inner and outer borders as indicated in the first row below the target name.
The minor axes $r_x \cos i$ and orientation of the ellipses follow from $i$ and the PA of the disks given in Table~\ref{tab: sample}. The relative uncertainties for the integrated count rates $\Delta Q_\varphi/Q_\varphi$ are estimated from the standard deviation of $Q_\varphi$ values derived for individual cycles. Typically $\Delta\hat{Q}_\varphi/\hat{Q}_\varphi \approx 10~\%$ for individual filter measurements with a few cases between $10~\%$ and $15~\%$. 
The precision is often better $\Delta\hat{Q}_\varphi/\hat{Q}_\varphi < 5~\%$ for the disk observations carried out with observing procedure A using simultaneous 
measurements of the disk polarization and the stellar intensity PSF.

The intrinsic disk polarization intensity $\hat{Q}_\varphi$ is obtained after correction of the $Q_\varphi(x,y)$ disk image with a $f(x,y)$ correction map as described in Sect.~\ref{sect:corr-map}. The PSF selection for the individual dataset is indicated in the last column. The derived intrinsic values for $\hat{Q}_\varphi$ have for measurements corrected with procedure A or C often smaller relative uncertainties than the initial ${Q}_\varphi$ values. 
This happens because the $f(x,y)$-corrections were applied to individual polarimetric cycles or different groups of cycles and this compensates also temporal variations of the PSF smearing effect present in the observational data. 

The intrinsic polarization $\hat{Q}_\varphi$ are typically a factor of about 1.2 to 2.3 higher than the measured values $Q_\varphi$ with a mean of about 1.6. The corrections are small for very extended disks, like HD~142527, and large for compact disks observed with low-quality instrumental PSF. The corrections are also systematically larger for visible wavelengths when compared to near-IR wavelength as can be easily inferred from the difference between observed $Q_\varphi/I_\star$ and intrinsic $\hat{Q}_\varphi/I_\star$. This systematic trend is expected for PSFs obtained with AO systems because the atmospheric turbulence introduces much larger wavefront distortions $\Delta\lambda/\lambda$ at shorter wavelengths. Therefore, the PSF correction applied in this work is so important to avoid strong, wavelength-dependent bias effects in the polarimetric measurements.

\begin{figure}
    \centering
    \includegraphics[width=0.48\textwidth]{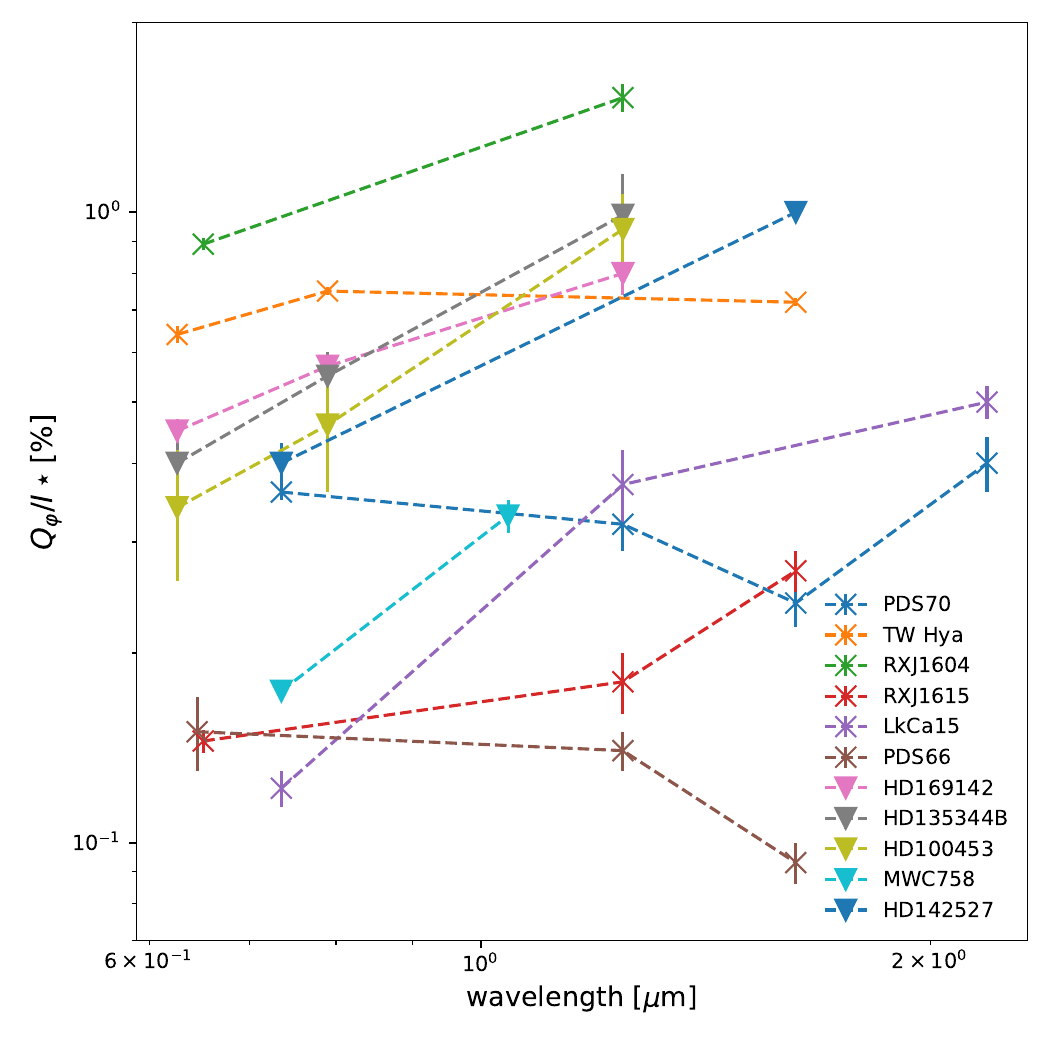}
    \caption{Convolution corrected or intrinsic polarized reflectivities $\hat{Q}_{\varphi}/I_\star$ for our T-Tauri star disks ($\times$) and Herbig star disks ($\triangledown$) measurements. For a given disk the results for different wavelength bands are connected with lines.}
    \label{fig:color_noscale}
\end{figure}

The most important parameters for the scientific interpretation are the derived polarized disk reflectivities $\hat{Q}_\varphi/I_\star$ or the intrinsic polarized intensities relative to the stellar intensity because these values are corrected for all observational effects. The obtained values $\hat{Q}_\varphi/I_\star$ are also shown in Fig.~\ref{fig:color_noscale}. 

What exactly does the $\hat{Q}_\varphi/I_\star$ values represent? The used $I_\star$ intensity overestimates the intrinsic flux of the star because $I_\star$ includes also the intensity signal of the extended disk. 
We estimate that the extended disk $I_{\rm disk}/I_\star$ is less than about 2~\% for about half of our measurements and between 2~\% and 5~\% for the other half. 
This assumes that the disk intensity is about four times larger than the measured polarized intensity $I_{\rm disk} \approx 4\cdot \hat{Q}_\varphi$ or that the disk averaged fractional polarization is about 25~\%.
However, using this "disk-contaminated" PSF for the determination of the correction map will slightly over-correct the smearing and therefore also over-estimate the calculated intrinsic  $\hat{Q}_\varphi$ for the disk. The small over-estimation of the star intensity $I_\star$ and the small over-estimation of the convolution will compensate each other and therefore the derived values 
$\hat{Q}_\varphi/I_\star$ represent good approximations for the ratios between the total polarized intensity of the disk and stellar intensity (unresolved central source) without the contribution of the extended disk. This approximation was tested in \citet{Ma2023} and confirmed to be very accurate for the disk around RX~J1604 using calculations for the PSF convolution of the best fitting disk model.

We find for our sample of bright disks values in the range $\hat{Q}_\varphi/I_\star\approx 0.1~\%$ to 1.5~\%. This relatively large spread is caused by the geometry of the disk and the accessible disk region for the intensity integration.
A thick or flared disk will intercept and scatter more light from the central star and produce a stronger scattering signal than a thin or flat disk. If the disk has a large central cavity, then all the scattering will take place at a large separation and be included in the integration. In contrast to this, a disk with no or only a small inner cavity will produce a large fraction of the scattered light at a separation that cannot be resolved observational, and then only the fraction of scattered light produced in regions further out can be measured.

\subsection{Notes on special cases}
\label{sect: special}
There are three targets in our sample, HD~100453, LkCa~15, and
PDS~70, with relatively high uncertainties for $\hat{Q}_\varphi/I_\star$ according to Table~\ref{tab:integrate-flux}
and therefore we discuss these cases individually.

\paragraph{HD~100453.} 
This is a bright disk and the $Q_\varphi$ values in the $J$ band from ten cycles of coronagraphic imaging polarimetry are very consistent. However, we find a discrepancy of 10~\% between the two available $I_\star$ observations taken in the same instrument configuration, which we cannot explain but could be caused by clouds. 
This introduces therefore a relatively large uncertainty for the derived $Q_{\varphi}/I_{\star}$ value. Also the $I_\star$ measurements for the $R'$ and $I'$ filters have relatively large uncertainties because the central star data are underexposed and the images for the disk polarimetry are strongly saturated at the location of the star. Higher precision measurements are achievable and should be obtained with a better flux calibration strategy.

\paragraph{LkCa~15.}
This is a faint target for AO observations in the visible region with SPHERE.
The inner disk is only partly resolved and we measure the polarization signal from the outer disk. The VBB data are not saturated and the measured brightness of the central star shows during the first run an intensity decrease of 25~\% and in the second run a constant intensity at the low level of the first run. 
LkCa~15 is an AA Tau-like variable star \citep{Alencar2018} and the observed brightness change could be a result of the variability of the central star. 

However, as the polarized disk intensity $Q_\varphi$ decreases at the same level while the relative polarization $Q_{\varphi}/I_{\star}$ stays constant, the variability is more likely caused by atmospheric transmission variations (thin clouds). 
The unsaturated PSF allows us to correct the polarized flux cycle by cycle, resulting in very small uncertainty $\sigma(\hat{Q}_{\varphi}/I_{\star})=\pm5\%$, despite the indicated large variability and uncertainty in $Q_\varphi$ and $I_\star$ in the VBB band.

\paragraph{PDS~70.}
Measuring uncertainties for PDS~70 are introduced by the strong polarization of the star and the unresolved inner disk as already reported by \cite{Keppler18}. We measure in an aperture with a radius of $0.072\arcsec$, after correction for the instrumental polarization, the values $p_{\star}=1.15\pm 0.12 \%$ and $\theta_{\star}=74 \pm 4^\circ$ for the VBB band, $0.90\pm 0.20~\%$ and $66 \pm 2^\circ$ for the $J$ band, and $1.1\pm 0.1~\%$ and $71\pm 4^\circ$ for the H band. This is not compatible with the expected polarization decrease of about 50~\% from VBB to H band for interstellar polarization indicating that the signal of the central source includes a substantial intrinsic polarization component. 
This signal is produced inside the integration range defined for the disk. The enhanced dispersion in the polarimetric intensity for this disk could be the result of a variable spillover of signal into the disk integration region from the strongly polarized central star. 
A more detailed analysis, considering each cycle individually, could be helpful to disentangle the different polarization components and obtain smaller measuring uncertainties. 

\subsection{Comparison with previous measurements}
\label{Sect:Comparison}
It is useful to check our results with previous disk measurements of the polarized intensity ${Q}_\varphi/I_\star$ and determinations of the corresponding intrinsic values $\hat{Q}_\varphi/I_\star$. 
The agreement of our $\hat{Q}_\varphi/I_\star$ values are within 10~\% when compared to previous values for HD~169142 \citep{Tschudi2021}, and RX~J1604 \citep{Ma2022}, probably because we used the same data and roughly the same integration region as these previous studies. 
For HD~142527, we obtain $\hat{Q}_\varphi/I_\star$ values which are about 13~\% and 7~\% smaller than \citet{Hunziker2021}. This is a rather large uncertainty that could be caused by the use of different background subtraction procedures.

The results for HD~135344B can be compared with the study of \cite{Stolker2016}. We measure $Q_\varphi/I_\star$ values which agree for the $J$ band but our values are about 33~\% lower for the $R'$ and $I'$ band. The reason for this discrepancy could be related to the larger flux integration regions for $Q_\varphi$ and $I_\star$ in \cite{Stolker2016}.

There also exists quite some discrepancy for the disk of LkCa15, for which \cite{Thalmann2016} measured $Q_{\varphi}/I_{\star}= 0.055\%$ for the VBB band and $0.36\%$ for the $J$ band. The result for the visible band marginally agrees with our VBB-band measurement (0.065~\%$)$, while our $J$ band value is only about ${Q}_{\varphi}/I_\star=0.25\%$ or about 70~\% of their value. 
Such discrepancies are difficult to explain because the measured $Q_\varphi/I_\star$ values depend strongly on temporal changes of the AO performance. Therefore, different $Q_\varphi/I_\star$ values can result, if different subsamples of the polarimetric disk images are selected, for example, if low-quality data are excluded.  Comparisons of $Q_\varphi/I_\star$ measurements without correction of the PSF smearing must therefore be considered carefully because the measuring results depend on the PSF quality of the selected observations.

A large difference exists between our work and the study of \citet{Avenhaus2018}, who derived also $\hat{Q}_{\varphi}/I_{\star}$ values for PDS~66 and RX~J1605 for the J and H band which are about a factor of three higher than our values because they included in their disk integration for these two disks the innermost, bright disk regions. We excluded the innermost regions for the disk color measurements because the quality of the used visible data is not good enough to resolve and include the innermost disk rings for PDS~66 and RX~J1605.

Many values for $Q_\varphi/I_\star$ of protoplanetary disks can be found in the literature. For example, the compilations of \citet{Garufi2020} and \citet{Garufi2022} give an overview of the measured $Q_\varphi/I_\star$ values for large samples. These studies are not focused on accurate photo-polarimetric measurements for the detailed characterization of the dust in the disks. Therefore, a comparison with these studies is difficult, because the values are not corrected for PSF smearing effects and the disk integration ranges are not accurately described. For several disks, our $Q_\varphi/I_\star$ results agree quite well (within a factor of two) with the compiled values but for several cases, we get much smaller values, probably because we used smaller disk integration regions like in the case of PDS~66 and RX~J1605 discussed above.

\paragraph{Convolution correction.} A quality assessment can also be made for the convolution correction procedure with the $f(x,y)$-maps introduced in this work with a comparison with the studies of \citet{Tschudi2021} for HD~169142 and \citet{Ma2023} for RX~J1604. In their studies, a parametric model is applied using forward modeling and PSF convolution. 
The good agreement between our results using the generalized correction map and the forward modeling was expected because these disks have a symmetric narrow ring geometry.

For the very extended disk HD~142527, the PSF smearing effects were corrected by \citet{Hunziker2021} with a single correction value derived with convolution calculation for an azimuthally symmetric disk ring. 
They estimated correction factors of 1.30 and 1.17 for VBB- and $H$ band which are in agreement with the flux-weighted correction factors 1.27 and 1.18, respectively, derived in this work based on the correction map $f(x,y)$. The approach of \citet{Hunziker2021} works well for an extended disk (like HD~142527 with a disk radius of $0.7''-1.4''$), where the smearing effects depend only weakly on radius. 
These three examples provide further support that the $f(x,y)$-map procedure yields accurate corrections for the PSF convolution effects for the derivation of the intrinsic disk polarization $\hat{Q}_\varphi/I_\star$ of transition disks.

\begin{table*}[]
    \centering
    \caption{Logarithmic wavelength gradients $\eta$ or colors for the polarized reflectivity for the measured disks.}
    \begin{tabular}{c | c c c}
    \hline
    \hline
    Host star  & $\eta_V$                         & $\eta_{V/IR}$                 & $\eta_{IR}$ \\
    \hline
    PDS~70     &                                  & $\eta_{VBB/J}= -0.22 \pm 0.23$ & $\eta_{J/H} = -1.08 \pm 0.66$ \\
               &                                  &                               & $\eta_{H/Ks} = 1.73 \pm 0.62 $ \\
    TW~Hya     & $\eta_{R'/I'} = 0.68 \pm 0.19$   & $\eta_{I'/H} = -0.12 \pm 0.12$ \\
    RXJ 1604   &                                  & $\eta_{R_d/J} = 0.83 \pm 0.12 $ \\
    RXJ 1615   &                                  & $\eta_{R_d/J} = 0.33 \pm 0.24$ & $\eta_{J/H} = 1.52 \pm 0.70$ \\
    LkCa~15    &                                  & $\eta_{VBB/J}= 2.11 \pm 0.38$  & $\eta_{J/Ks} = 0.54 \pm 0.35$ \\
    PDS~66     &                                  & $\eta_{N_R/J} =-0.11 \pm 0.31$ & $\eta_{J/H} = -1.54 \pm 0.55$ \\
    HD~169142  & $\eta_{R'/I'} = 1.02 \pm 0.34$   & $\eta_{I'/J} = 0.74 \pm 0.24$ & \\
    HD~135344B & $\eta_{R'/I'} = 1.37 \pm 0.72$   & $\eta_{I'/J} = 1.29 \pm 0.55$ & \\
    HD~100453  & $\eta_{R'/I'} = 1.30 \pm 1.95$   & $\eta_{I'/J} = 1.57 \pm 0.78$ & \\
    MWC~758    &                                  & $\eta_{VBB/Y} = 1.83 \pm 0.24$ & \\
    HD~142527  &                                  & $\eta_{VBB/H} = 1.16 \pm 0.10$ & \\
    \hline
    \end{tabular}
    \label{tab:color-gradient}
\end{table*}

\section{Color of the polarized reflectivity}
\label{sect:colors}
All our determinations for the disk integrated polarized reflectivity $\hat{Q}_\varphi/I_\star$ are shown in Fig.~\ref{fig:color_noscale} as a function of wavelength including lines connecting for a given target the values for different $\lambda$. 
One can expect that these wavelength differences between the polarized signal $\hat{Q}_\varphi/I_\star$ are mainly caused by scattering and absorption properties of the dust in the disk because we can assume that the scattering geometry for a given disk is very similar for the different wavelengths. Therefore, the wavelength dependence $(\hat{Q}_\varphi/I_\star)_\lambda$ is an excellent source of information for the investigation of dust properties. 

The lines in Fig.~\ref{fig:color_noscale} indicate that $\hat{Q}_\varphi/I_\star$ increases clearly with wavelength for one group of disks (HD~100453, HD~135334B, HD~142527, HD~169142, MWC~758, LkCa~15, RX~J1604) while another group, consisting of TW~Hya, RX~J1615, PDS~66, and PDS~70, shows rather similar values for different wavelengths. 

We use the logarithmic wavelength gradient of the polarized
reflectivity between two wavelengths $\lambda_1<\lambda_2$ as a measure of the color \citep{Tazaki2019}, 
\begin{equation}
    \eta_{\lambda_1/\lambda_2} = \dfrac{\log(\hat{Q}_{\varphi}/I_{\star})_{\lambda_2}-\log(\hat{Q}_{\varphi}/I_{\star})_{\lambda_1}}{\log(\lambda_2/\lambda_1)}\ ,
\end{equation}
which are given in Table~\ref{tab:color-gradient} for many combinations of wavelength bands. 
We introduce three dust color classes for the disks, according to the measured reflectivity gradient: blue for $\eta<-0.5$, gray for $-0.5<\eta<0.5$, and red for $\eta>0.5$.
A red color $\eta=1$ is equivalent to the magnitude colors of 
$m_{R'}-m_{I'}=0.25^m$, $m_{I'}-m_{J}=0.49^m$, or $m_{J}-m_{H}=0.29^m$.

\begin{figure}
    \centering
    \includegraphics[width=0.48\textwidth]{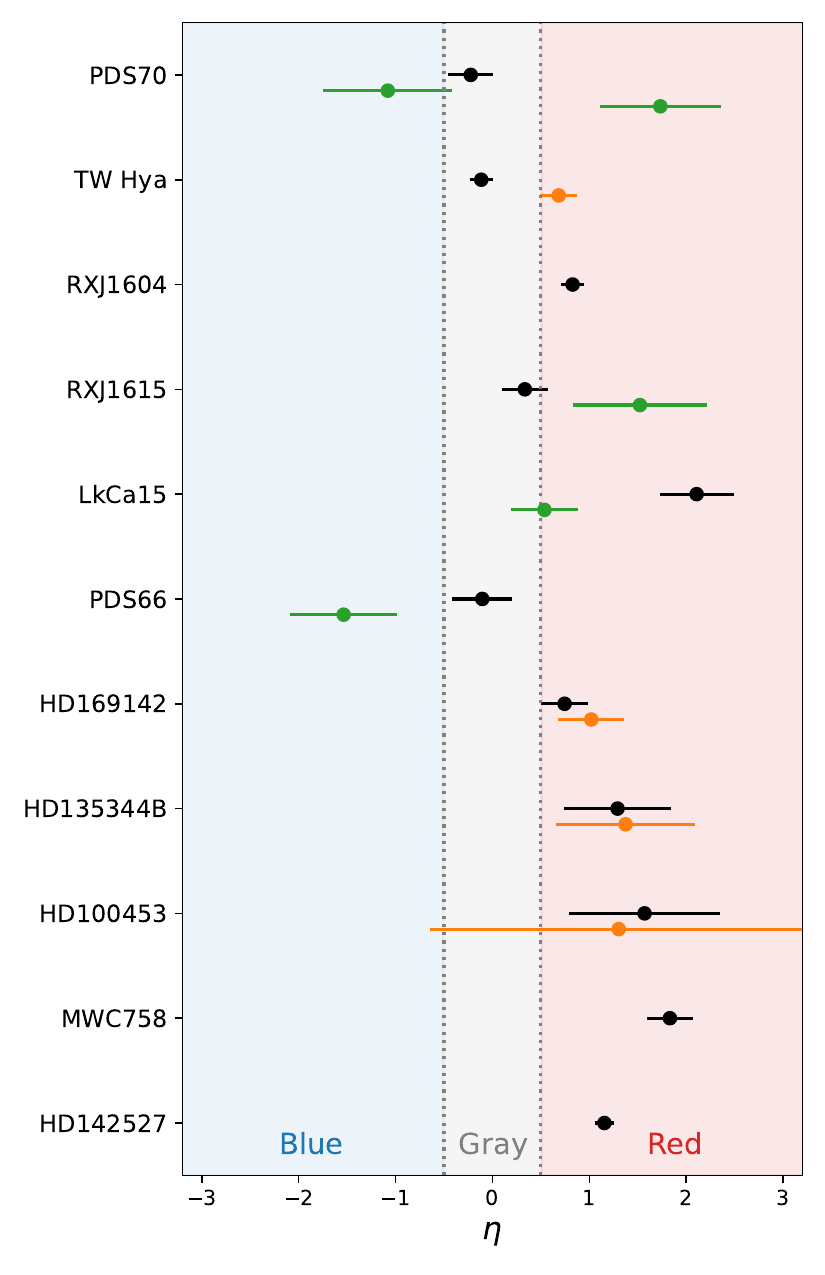}
    \caption{Logarithmic wavelength gradients $\eta$ for the polarized reflectivity for the measured disks. Targets are sorted by the stellar luminosity. Black points are colors $\eta_{\rm V/IR}$ between visible and near-IR bands, orange points for visible bands $\eta_{\rm V}$, and
    green points for near-IR bands $\eta_{\rm IR}$. The shading represents the
    used definition of blue, gray, and red disk reflectivities.}
    \label{fig:color-gradient}
\end{figure}

We compute the colors for each target between two consecutive wavelength bands and distinguish between three wavelength types of colors: colors $\eta_{V/IR}$ derived from visible $\lambda_1 < 1~\mu$m and near-IR $\lambda_2> 1~\mu$m wavelengths, colors $\eta_{V}$ based on visible wavelengths only ($\lambda_1,\lambda_2<1~\mu$m), and colors $\eta_{IR}$ based on near-IR wavelengths only ($\lambda_1,\lambda_2>1~\mu$m). 
All targets have been measured at visible and near-IR wavelengths, but the wavelengths for visible or near-IR bands are not always the same. For example, for HD~135344B we derive the color $\eta_{V/IR} = \eta_{I/J}$ and for MWC~758 the color $\eta_{V/IR} = \eta_{VBB/Y}$. The visible color is $\eta_{V} = \eta_{R'/I'}$, because $R'$ and $I'$ can be measured simultaneously in the two arms of ZIMPOL. The near-IR color is for three cases $\eta_{IR}=\eta_{J/H}$, but there are also colors based on $\eta_{J/Ks}$ for LkCa~15 and $\eta_{H/Ks}$ for PDS~70. The results are shown in Fig.~\ref{fig:color-gradient}, with $\eta_{V}$, $\eta_{V/IR}$, and $\eta_{IR}$ plotted in orange, blue, and green respectively. 

All our disk gradients fall roughly into the range $-1.5<\eta<2$. 
The colors for the polarized reflectivity seem to show a trend for the two subgroups of our sample, with a lower parameter $\eta_{V/IR}$ for the T-Tauri disks when compared to the Herbig star disks. All disks around Herbig stars are red and have $\eta_{V/IR}>0.5$, while four out of six disks around T-Tauri stars (PDS~70, TW~Hya, RX~J1615, and PDS~66) have gray colors with $-0.5< \eta_{V/IR}<0.5$. The disk around the T-Tauri star RX~J1604 is "slightly" red ($\eta_{V/IR} \approx 0.83$) while the disk around the T-Tauri object LkCa~15 is an outlier with very red colors ($\eta_{V/IR} \approx 2.11$).

\subsection{Comparison with previous color determinations}
Our measurements of the polarized reflectivities $\hat{Q}_\varphi/I_\star$ for HD~142527, HD~169142, RX~J1604, and HD~135334B agree well with previous studies \citep{Hunziker2021,Tschudi2021,Ma2023,Stolker2016} as discussed in Sect.~\ref{Sect:Comparison}. 
For these objects, the corresponding polarized reflectivity colors $\eta$ given in Tab~\ref{tab:color-gradient} and Fig.~\ref{fig:color-gradient} also confirm the previous color determination and provide for HD~169142 the first $\eta_{\rm V/IR}$ value and for HD~135344B improved color values with error bars.

The disk of PDS~66 is interesting because we measure a blue color $\eta_{J/H}=-1.54$ in the infrared. A blue J/H color for this target was also noticed by \citet{Avenhaus2018}. They indicate large or conservative uncertainties for their measurements and therefore did not obtain a significant blue color for the polarized reflectivity. 
When disregarding the uncertainties, PDS~66 is the bluest disk in the sample of \citet{Avenhaus2018} with a $Q_\varphi/I_\star$ ratio between the $J$ band and $H$ band of 1.27, which corresponds to a logarithmic color gradient of $\eta_{J/H}=-0.90$.
The other target also included in \citet{Avenhaus2018} is RX~J1615 where they get a $J/H$ band ratio of $0.78$, or a color gradient of $\eta_{J/H}=0.95$.
It should be noted that their measurements include the innermost bright rings of these two targets (Fig.~\ref{fig:gallery}), resulting in about three times larger $\hat{Q}_\varphi/I_\star$ values. Nevertheless, the derived colors for the polarized reflectivities agree quite well with the determination of PDS~66 and RX~J1605 in \citet{Avenhaus2018} within the errorbars.
The agreement for the reflectivity colors $\eta$ could be the result of the fact that the same measuring procedure for $\hat{Q}_\varphi/I_\star$ was used for the $J$ and $H$ band in the individual studies, and this compensates probably for systematic measuring offsets in an appropriated way in the differential determination of the color $\eta$.

Our polarimetric color measurements $(\hat{Q}_\varphi/I_\star)_\lambda$ of TW~Hya can be compared to the multiwavelengths HST measurements of the scattered intensity $(I_{\rm disk}/I_\star)_\lambda$ presented in \citet{Debes2013}. They find roughly gray colors for the reflected intensity consistent with the colors derived for the polarized reflectivity $(\hat{Q}_\varphi/I_\star)_\lambda$ in this work. These results can be combined to constrain the wavelength dependence of the fractional polarization $p_{\rm disk}(\lambda)$ of TW~Hya, and we estimate, that $p_{\rm disk}(\lambda)$ is roughly constant from the $R$ band to the $H$ band within an uncertainty band of about $\pm 30~\%$. This excludes a strongly increasing fractional polarization with wavelengths $p_{\rm disk}(\lambda)$ as predicted with some dust aggregate models with a single monomers size of 100 nm
\citep[e.g.,][]{Tazaki2022}.

The comparison of our dust color measurements and previous studies reveals no significant discrepancies. However, our work provides for nine targets new or much improved values for the color $\eta$ of the polarized reflectivity and provides now typical or mean values for the polarized colors $\eta$ a large sample and a corresponding range of measured color which is characteristic for transition disks.  

\subsection{Search for radial color variations}
\label{Sect:in-out}
The disk measurements show significant differences for the colors of the polarized reflectivity between different disk targets (Fig.~\ref{fig:color-gradient}). It is therefore also interesting to investigate, whether there exist color differences for the polarized scattered light within a given disk, for example between the inner disk region and the outer disk region. Our disk sample and our measuring procedure using elliptical annuli is not optimized for the search of radial color variations. But three disks in our sample, HD~169142, HD~135344B, and HD~100453 can be measured by an inner and an outer annulus as indicated by the dashed ellipse in Fig.~\ref{fig:gallery}.
HD~145344B and HD~100453 have extended spirals that can be separated from the inner ring and HD~169142 has two well-defined rings. Table~\ref{tab:substructure} lists polarized intensity determinations $\hat{Q}_\varphi/I_\star$ for the inner ring and the outer structures for these disks for all available wavelength bands. For a given disk the flux ratios between the inner ring and outer structures are the same within the measuring uncertainties for different bands. This indicates that there are no significant color differences between the reflectivities of the inner and outer disk structures.  

\begin{table}[]
    \centering
    \caption{Integrated polarized intensity relative to the intensity of the central source, $\hat{Q}_\varphi/I_\star$ in [\%], for the inner ring and the outer disk structures for three objects.}
    \label{tab:substructure}
    \begin{tabular}{c|c c c c}
    \hline
    \hline
    Host star     & Filt   & Inner             & Outer              & Ratio \tablefootmark{a} \\
                  &        & ring              & structures         & \\
    \hline
    HD~135344B    & R'     & 0.20 $\pm$ 0.02   & 0.20 $\pm$ 0.01    & 1.0 $\pm$ 0.1\\
    $[0.10,0.26]$ & I'     & 0.28 $\pm$ 0.03   & 0.27 $\pm$ 0.02    & 1.0 $\pm$ 0.2\\
    $[0.26,1.15]$ & J      & 0.46 $\pm$ 0.09   & 0.53 $\pm$ 0.07    & 0.9 $\pm$ 0.3\\[0.12cm]
    HD~100453     & R'     & 0.23 $\pm$ 0.06   & 0.11 $\pm$ 0.02    & 2.1 $\pm$ 0.9\\
    $[0.10,0.23]$ & I'     & 0.31 $\pm$ 0.07   & 0.16 $\pm$ 0.03    & 1.9 $\pm$ 0.8\\
    $[0.23,0.86]$ & J      & 0.64 $\pm$ 0.09   & 0.30 $\pm$ 0.05    & 2.1 $\pm$ 0.6\\ [0.12cm]
    HD~169142     & R'     & 0.30 $\pm$ 0.01   & 0.14 $\pm$ 0.01    & 2.1 $\pm$ 0.2\\
    $[0.10,0.36]$ & I'     & 0.40 $\pm$ 0.02   & 0.17 $\pm$ 0.02    & 2.4 $\pm$ 0.4\\
    $[0.36,1.08]$ & J      & 0.59 $\pm$ 0.04   & 0.21 $\pm$ 0.01    & 2.8 $\pm$ 0.3\\
    \hline
    \end{tabular}
    \tablefoot{The integration region is given below the target name by [r1,r2] in arcsec for the inner ring and outer structures respectively. \tablefoottext{a}{Flux ratio between the inner ring and the outer structures}}
\end{table}

Also, TW~Hya, RX~J1615, and PDS~66 show clearly separated inner and outer disk structures. But the inner working angle of the polarized imaging data does not allow accurate color measurements for small separations and therefore the inner disk was excluded for the polarized flux integration given in Table~\ref{tab:integrate-flux}.
A special case is the disk around MWC~758 with its inner ring structure and the two extended spirals, but the ring and spirals cannot be separated properly with an ellipse (inclined ring). However, it seems that there exists a color difference between the northern and the southern arm. A special measuring procedure is therefore required to quantify the color differences between substructures of MWC~758 and we defer this to a future study.

\section{Interpretation of the reflectivity colors}
\label{sect:interpretations}
\begin{figure}
    \centering
    \includegraphics[width=0.45\textwidth]{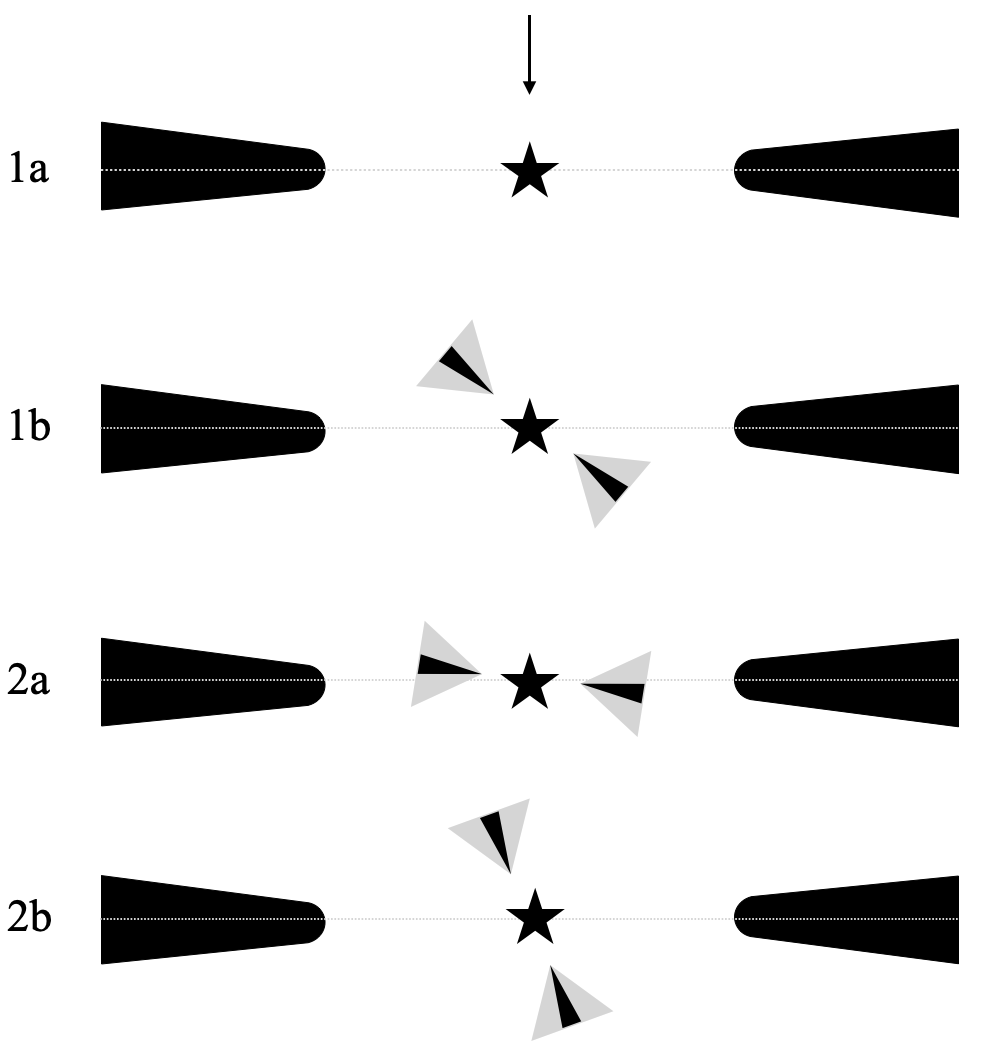}
    \caption{Sketches of different disk geometries that may affect the measured disk color. The arrow indicates the view direction. The dotted lines indicate the mid-plane in each case. The black and gray wedges represent the optically thick and optically thin layer of the inner disk respectively.}
    \label{fig:scenario}
\end{figure}

In this study, we derive the wavelength dependency or color of the polarized reflectivity $(\hat{Q}_\varphi/I_\star)_\lambda$ for 11 disks with the goal to obtain additional constraints on the properties of the scattering dust. In addition, we search for systematic differences between different types of disks which may point to dust evolution processes in these systems. We consider two basic scenarios for dust scattering for the interpretation of the derived colors and illustrate the geometric scenarios in Fig.~\ref{fig:scenario}.

In scenario 1, the disk reflectivity color is predominantly caused by the wavelength dependence of the dust scattering in the disk region selected by our $Q_\varphi$-measurements. Scenario 1 allows a direct determination of the scattering properties of the dust, because possible effects of the inner hot disk, or hot diffuse dust near the star can be neglected. This scenario probably applies, if there is no or only weak thermal emission from hot dust in the near-IR as shown for the 1a case in Fig.~\ref{fig:scenario}. 
Alternatively as shown in the 1b case, a geometrically thin but tilted inner disk may cast only narrow shadows on the outer disk without changing significantly the spectral energy distribution (SED) of the stellar illumination. 

In scenario 2, the color of the outer disk is a result of two effects, the dust reflection in the observed disk region and the wavelength-dependent absorption, emission, and scattering caused by hot dust near the central star. For example, in case 2a, optically thin absorption near the star by small particles elevated above the disk midplane produces a reddening of the stellar radiation and this may change significantly the SED of the radiation illuminating the disk. This is shown in Fig.~\ref{fig:scenario} case 2a for a slightly tilted inner disk, but an aligned inner disk could produce the same effect. Alternatively, in case 2b the dust may cause absorption for "our" line of sight to the star, and change the SED of the stellar spectrum. This happens for example for a photometric minimum in the AA Tau-like variable stars RX J1604 and LkCa 15. This is considered in our analysis for these two objects, but such an effect might remain unrecognized, if it is permanent and does not produce variability.
The 2a case would introduce a red offset and the 2b case a blue offset from the true color for the derived $\eta$ value of the disk reflectivity.  

Another problem in scenario 2 is the thermal emission of the hot dust,
which can contribute to the flux of the unresolved central source in the $Ks$ band (2.2~$\mu$m) or even in the $H$ band (1.6~$\mu$m) while shorter wavelengths are much less or not affected. 
If the hot dust is not emitting isotropically because of the disk structure, it might introduce another positive or negative differential effect between the SED of the directly observed central source and the SED for the radiation illuminating the disk. For our disk sample, none of the measurements should be strongly affected because we did not use $Ks$ and $H$ band data for the targets with strong near-IR emission, namely HD~169142, HD~135334B, HD~100453, and MWC~758 (Table~\ref{tab: sample}).

Detailed studies of each target would be required to clarify, whether scenario 1 or 2 applies and whether corrections must be applied to the derived colors to characterize the scattering dust in the disk. Which scenario applies, depends on the detailed disk geometry and our viewing angle to the system.
For example, the 2b case degenerates to the 1b case if the disk is seen close to edge-on. 
It is beyond the scope of this work to investigate these aspects in detail for each target, but we subsequently consider the possible impact of scenario 2 for the interpretation of our color determinations.

\subsection{Disk colors and dust particle properties}
We shall first assume scenario 1 applies and that the observed reflectivity color is only caused by the wavelength dependency of the dust scattering properties. In this case, we can base our discussion on the results for the simplified, parametric transition disk models of \citet{Ma2022}. 
The models provide polarized disk reflectivities $\hat{Q}_\varphi/I_\star$ for dust described by three parameters only, the single scattering albedo $\omega$, a Henyey-Greenstein scattering phase function for the intensity $f(\theta)$ with asymmetry parameter $g$, and a scaled Rayleigh scattering function for the polarization $p(\theta)$ with the maximum fractional polarization parameter $p_{\rm max}$. The disk geometry is described by the inclination $i$ of the disk mid-plane, the slope of the inner disk wall $\chi$, and an angular disk height $\alpha$.  

Within their model grid, \citet{Ma2022} define also a reference case called the "0.5-model" with dust parameters $\omega=0.5$, $g=0.5$ and $p_{\rm max}=0.5$.
This yields for $i=32.5^\circ$, $\chi=32.5$ and $\alpha=10^\circ$ the normalized disk polarization $\hat{Q}_\varphi/I_\star=0.69~\%$ similar to the values derived for the disks in Table~\ref{tab:integrate-flux}.
$\hat{Q}_\varphi/I_\star$ depends roughly linearly on the adopted disk height $\alpha$ and is therefore not constraining well the dust scattering parameter. However, the averaged fractional polarization of the "0.5-model" is $\hat{Q}_\varphi/I_{\rm disk}=\langle p_\varphi \rangle=24~\%$. This value depends mostly on the dust scattering parameters and is almost independent of the disk parameters $\alpha$, $i$, and $\chi$ for low inclinations ($i<57.5^\circ$) and not too steep wall slopes ($\chi<57.5^\circ$) appropriate for the investigated disks.
Measurements, yield values in the same range $\langle p_\varphi \rangle \approx 20~\%$ to $45~\%$ \citep{Hunziker2021, Tschudi2021, Ma2023} indicating that the scattering parameters of the "0.5-model" are a reasonable approximation or starting point for our discussion. 
Based on this, the observed colors of the polarized reflectivity $(\hat{Q}_\varphi/I_\star)_\lambda$ and the results of \citet[][Fig.~10 and Tab.~2]{Ma2022} near the reference scattering parameters ($\omega,g,p_{\rm max}\approx 0.5$) wavelength dependencies which can be approximated by
\begin{equation}
\frac{{\rm d}(\hat{Q}_\varphi/I_\star)_\lambda}{{\rm d}\lambda}\approx 
          1.1 \, \frac{{\rm d}\omega(\lambda)}{{\rm d}\lambda} - 0.8\, \frac{{\rm d}g(\lambda)}{{\rm d}\lambda} 
              + 1.0 \,\frac{{\rm d}p_{\rm max}(\lambda)}{{\rm d}\lambda}\,. 
\label{Eq.refpoint}            
\end{equation}
A gray color ${\rm d}(\hat{Q}_\varphi/I_\star)_\lambda/{\rm d}\lambda\approx$ 0 means that the scattering parameters do not change with wavelength or that the wavelength dependencies described in Eq.~(\ref{Eq.refpoint}) compensate each other. 

\begin{figure*}[!h]
    \centering
    \includegraphics[width=0.8\textwidth]{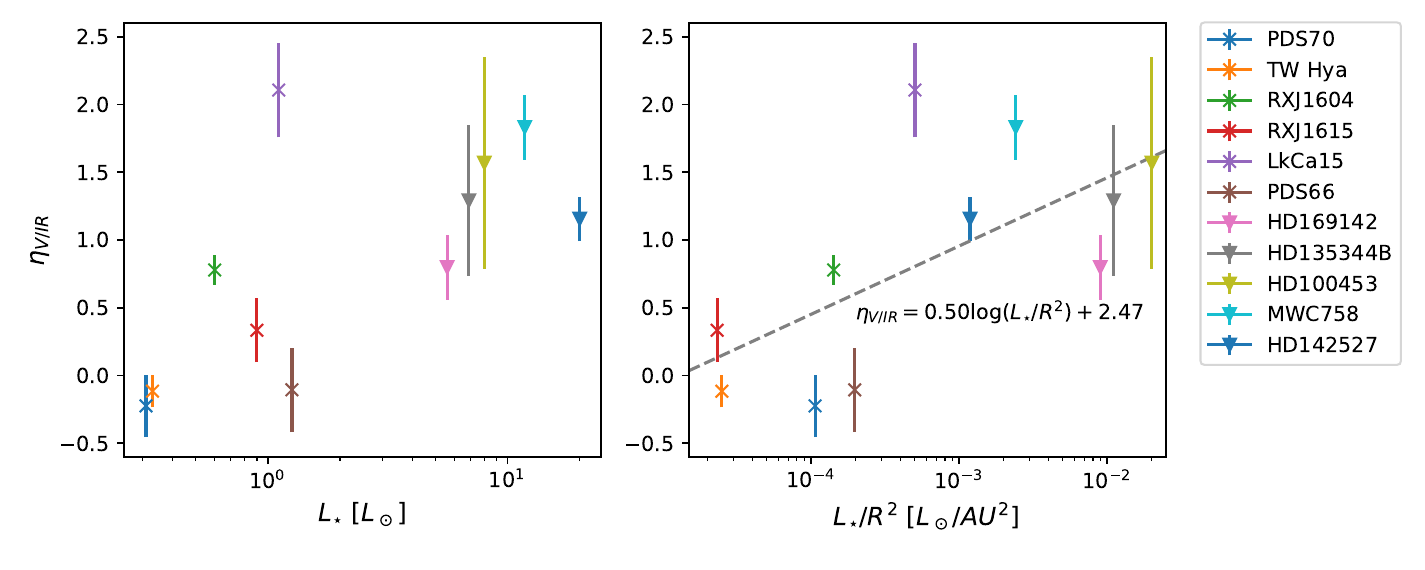}
    \caption{Correlation analysis of the disk color gradient and system parameters. Left: Correlation between the color gradient of the polarized reflectivity $\eta_{\rm V/IR}$ and stellar luminosity $L_\star$. Right: Correlation between $\eta_{\rm V/IR}$ and effective stellar irradiation $L_\star/R^2$.}
    \label{fig:correlation-luminosity}
\end{figure*}

The red color derived for one part of our sample could be explained by $\omega(\lambda)$ or $p_{\rm max}(\lambda)$ increasing with $\lambda$, or $g(\lambda)$ decreasing with $\lambda$, or a combination of these effects. 
Equation~(\ref{Eq.refpoint}) sets strong constraints, because a color of $\eta_{V/IR} \approx 1$ is equivalent to a doubling of $\hat{Q}_\varphi/I_\star$ between the $R'$ and the $J$ band, or between the $I'$ and the $H$ band. This would for example require that $\omega$ increases from 0.5 to 0.8, $g$ decreases from 0.7 to 0.4, and $p_{\rm max}$ increases from 0.4 to 0.8 between the visible and the near-IR. This guess implies quite strong wavelength dependencies for the dust scattering parameters, but they are still at a reasonable level. 
\smallskip

The observed scattered light properties of protoplanetary disks can be compared with detailed dust model calculations to characterize the dust properties. 
Dust consisting mainly of small grains $a< 1~\mu$m, like interstellar dust, is compatible with $\omega$ and $g$ decreasing with $\lambda$ and $p_{\rm max}$ increasing with $\lambda$, between the visible and near-IR range \citep[e.g., ][]{Whitney1995, Draine2003}, because the single particle scattering approaches the Rayleigh regime for wavelengths $\lambda > 1~\mu$m. Therefore, this dust can not produce the observed red colors and the strongly forward scattering of protoplanetary disks because of the low $\omega$ and $g$ in the near-IR as pointed out by \citet{Mulders2013}. 
Larger ($a\approx 1~\mu$m) spherical grains \citep{Mulders2013} or compact aggregates \citep{Min2016} yield forward scattering dust and red dust reflectivities but the resulting fractional polarization is low when compared to disk observations. Disks with a high scattering polarization and strong forward scattering are obtained for porous dust aggregates composed of small monomers by \citet{Tazaki2019}. But these dust models yield mostly blue and gray disk colors and no red colors and the reflectivity of the disk is rather high $(I_{\rm disk}/I_\star \geq 10~\%$) when compared to observations. This comparison between models and observations may indicate that dust with $a\approx 1~\mu$m with a porosity intermediate between the models of \citet{Min2016} and \citet{Tazaki2019} or aggregates with a range of compositions and monomer sizes \citep{Tazaki2023} could provide a rather good match to the available color measurements.

For a detailed characterization of the dust in protoplanetary disks, one should not only consider the disk colors for the polarized reflectivity $(\hat{Q}_\varphi/I_\star)_\lambda$ as derived in this work. Additional constraints can be obtained for bright, extended disks by the combination of $(\hat{Q}_\varphi/I_\star)_\lambda$ and intensity $(I_{\rm disk}/I_\star)_\lambda$ measurements for the derivation of the fractional polarization of the reflected light $p_\varphi(\lambda)$. 
Further one should estimate the scattering albedo $\omega$ with a comparison of the amount of scattered light $I_{\rm disk}/I_\star$ or $\hat{Q}_\varphi/I_\star$ with the absorbed stellar radiation as derived from the infrared excess $F_{\rm IR}/F_\star$ as obtained for HD~142527, HD~169142 and RX~J1604 \citep{Hunziker2021, Tschudi2021, Ma2023}. For inclined disks, one can also constrain the asymmetry parameter $g(\lambda)$ with measurements of the brightness contrast between the front and back side. 

Of course, determinations for only a few objects may not be representative for the whole class of protoplanetary disks and investigations for a larger sample, as in this work for the $(\hat{Q}_\varphi/I_\star)_\lambda$, helps to define the typical range and average values of radiation parameters. 
Another study of a larger disk sample study is the compilation of the ratio between polarized reflectivity and infrared excess $\Lambda_\varphi=(Q_\varphi/I_\star)/(F_{\rm IR}/F_\star)$ by \citet{Garufi2018}. 
They obtain a typical value of about $\Lambda_\varphi\approx 3~\%$ for transition disks, which is compatible with $\omega = 0.5$ of the reference model \citep{Ma2022}.
Similarly, the measurements for the maximum fractional polarization ${\rm max}(p_{\rm disk})$ measured on disks collected by \citet{Tazaki2022} indicates that the typical polarization parameter $p_{\rm max}$ for dust scattering is $p_{\rm max} \gtrapprox 30~\%$ in the visible and is $p_{\rm max}\gtrapprox 45~\%$ in the near-IR. 
The dust scattering asymmetry $g$ has been investigated by \citet{Ginski2023}, who measured the polarized surface brightness of disks as a function of scattering angle $\theta$ and they found that many disks have dust producing strong forward scattering based on the flared disk model of \cite{Stolker2016b}. 
More such quantitative measurements for protoplanetary disks will become available and provide further constraints on the scattering dust.

In the discussion above, it is assumed that the measured colors for $(\hat{Q}_\varphi/I_\star)_\lambda$ are only caused by the wavelength dependencies of the dust scattering parameters $\omega(\lambda)$, $g(\lambda)$, and $p_{\rm max}(\lambda)$ according to scenario A or Eq.~(\ref{Eq.refpoint}). However, based on the existing dust models it seems very difficult to explain the extreme colors $\eta_{V/IR} \approx 2$ for LkCa~15 and MWC~758. 

Therefore one should consider for these cases also scenario B, where the SED of the radiation illuminating the observed disk is undergoing an "extra"-reddening by absorbing dust near the central star. This scenario was also proposed previously for the disk around GG Tau to explain the red color for the scattered intensity measured by HST \citep{Krist2005}. It is even possible that a small amount of such "extra"-reddening is a frequent phenomenon for protoplanetary disks. 

\subsection{Dust colors and system properties}
The apparent difference of the polarized reflectivity colors $\eta_{\rm V/IR}$ between predominantly gray disks around T-Tauri stars and red disks around Herbig stars noted in Sect.~\ref{sect:colors} may point to a systematic dependence of the dust properties on the parameters of the central star. Obvious differences between these two stellar types are the higher luminosity and the bluer SED of the Herbig stars. We searched therefore for correlations between the system parameters given in Table~\ref{tab: sample} and the reflectivity colors $\eta_{\rm V/IR}$ (Table.~\ref{tab:color-gradient}) for our disks and there seems to exist trends between $\eta_{\rm V/IR}$ and the "strength" of the stellar illumination of the disk. 

Figure~\ref{fig:correlation-luminosity} shows diagrams for $\eta_{\rm V/IR}$ and the luminosity log $L_\odot$ of the central star, and $\eta_{\rm V/IR}$ and the effective stellar illumination $\log (L_\odot/R^2)$ at the separation $R$ of the main scattering component. Similar diagrams result if $\eta_{\rm V/IR}$ is plotted with respect to stellar spectral types or surface temperature (not shown). The dashed regression line for $\eta_{\rm V/IR}$ vs. $\log (L_\odot/R^2)$ has a Pearson parameter of about $0.7$.
This corresponds for such a small sample to only a nonzero correlation probability of about 90~\% which is not sufficient to be called significant. 
We find no clear trends between color $\eta_{\rm V/IR}$ and age of the system, or $\eta_{\rm V/IR}$ and disk inclination, or $\eta_{\rm V/IR}$ and far-IR excess. 

\begin{figure}
    \centering
    \includegraphics[width=0.48\textwidth]{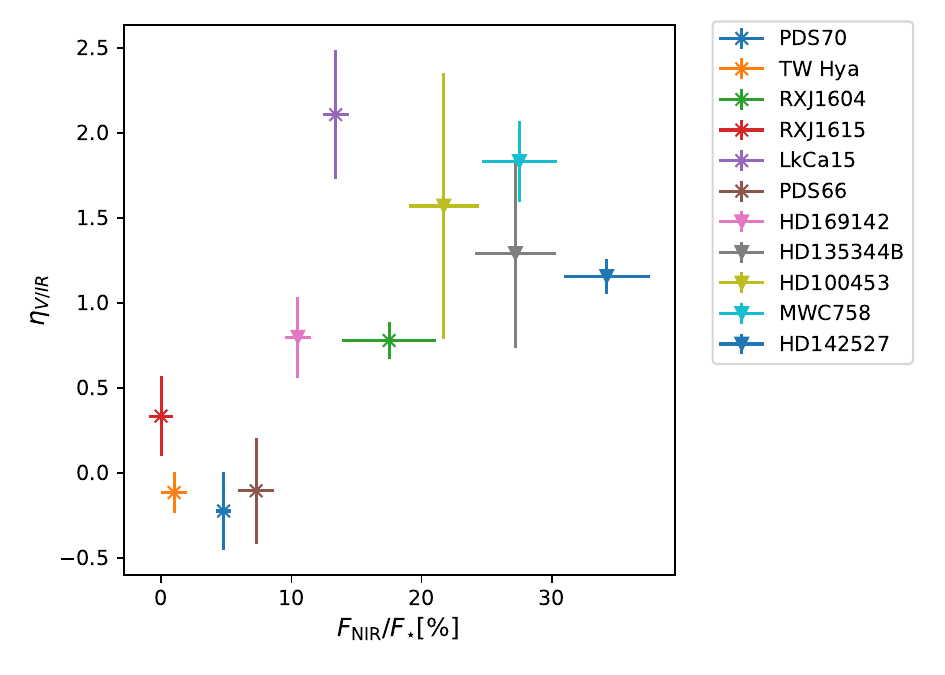}
    \caption{Correlation between the color gradient of the polarized reflectivity 
    $\eta_{\rm V/IR}$ and the near-IR excess for our disk sample. }
    \label{fig:correlation-NIR}
\end{figure}

However, there seems to exist also a trend between color $\eta_{\rm V/IR}$ and the near-IR excess $\log (F_{\rm NIR}/F_\star)$ as shown in Fig.~\ref{fig:correlation-NIR}, but also this trend is not significant. 
As discussed at the beginning of Sect.~\ref{sect:interpretations} for scenario 2, the hot dust responsible for the near-IR excess could produce by absorption an "extra"-reddening of the SED for the radiation illuminating the disk. 
Most likely, the "extra"-reddening requires optically thin dust in a geometric configuration so that the SED of the directly observed stellar light is less affected than the SED of the radiation illuminating the disk, like in case 2a. Under this special geometric configuration, the inner disk might become more puffed up due to the stronger irradiation from Herbig stars, which means wider gray wedges in Fig.~\ref{fig:scenario}. The more flared disk will increase the IR excess and will also intercept and redden more radiation that is traveling toward the outer disk, which results in the observed correlation between the disk color, the stellar irradiation, and the IR excess. 

This does not apply for systems with strongly inclined inner disk with
respect to the outer disk plane like case 1b and 2b in Fig.~\ref{fig:scenario}. 
These geometries produce well-defined shadows on the outer disk which is outside of the shadows bright and apparently strongly illuminated like for HD~142527 \citep{Marino2015}, HD~100453 \citep{Benisty2017}, or RX~J1604 \citep{Pinilla2018, Ma2023}. 
Detailed radiation models are required for a reliable interpretation of the measured colors for such special disks but this is beyond the scope of this study.

Thus, our sample shows seemingly but not significant trends between the color of the polarized disk reflectivity $\eta_{\rm V/IR}$ and
the stellar illumination but also between $\eta_{\rm V/IR}$ and the near-IR excess from hot dust. Both aspects should be further investigated with a larger disk sample because constraining and understanding such possible correlations would be very important for the correct interpretation of the colors.
Including targets having parameter combinations not covered in the sample presented in this work could be particularly useful. For instance, it would be beneficial to investigate disks around Herbig stars with little near-IR excess and disks around T-Tauri stars with large near-IR excess. 
In addition, a slightly misaligned system can cast a wide shadow on the outer disk as described by \citet{Muro-Arena2020} for HD~139614. The analysis of such a wide shadow system could be interesting to disentangle the effects caused by the inner disk and the dust scattering in the outer disk.

\section{Conclusions}
\label{sect:conclusions}
This work presents measurements of the polarized reflectivity $Q_\varphi/I_\star$ for a sample of 11 protoplanetary disks based on high-resolution imaging polarimetry using the SPHERE AO system. The data are corrected for the PSF smearing and polarimetric cancellation effects with a procedure that can be applied to a wide range of disk observations. It is demonstrated, that we obtain corrected or intrinsic values for $\hat{Q}_\varphi/I_\star$ with relative errors of only about 10~\% or even less. The achievable uncertainties depend mostly on the quality of the PSF calibration. The errors are particularly small for simultaneous measurements of the polarized intensity of the disk and the intensity PSF of the central star. 

Accurate $\hat{Q}_\varphi/I_\star$ values for different wavelengths are required to derive constraining wavelength dependencies or colors like $\eta_{\rm V/IR}$ for the disk reflectivities.
We derive colors $\eta_{\rm V/IR}$ which are significantly different between different targets as we find red colors $\eta_{\rm V/IR}>0.5$ for all disks around Herbig stars in our sample, while four out of six disks around T-Tauri stars show gray colors. 

Measuring the wavelength dependence or color of $(\hat{Q}_\varphi/I_\star)_\lambda$ increases significantly the diagnostic power of circumstellar disk observations. 
The disk models, which are already constrained by the apparent disk geometry and the measured IR excess, must now explain not only the amount of polarized light but also its wavelength dependence. Model parameter "knobs", which could be used previously to adjust the amount of polarized light in a given disk must now achieve this also for different wavelengths. This restricts strongly the dust scattering parameters for models capable to reproduce the multiwavelength reflectivity observations.

Unfortunately, there exist only a few studies that investigate the wavelength dependence of the scattered light of protoplanetary disks, and therefore the interpretation of the obtained color values $\eta$ is unclear. We find, based on the parametric study of \citet{Ma2022}, that the observed gray colors $\eta_{\rm V/IR}\approx 0$ or moderately red colors $\eta_{\rm V/IR}\approx +1$ are compatible with the dust having scattering parameters varying with wavelengths within the parameter ranges $\omega\approx 0.5-0.8$, $g\approx 0.3-0.8$ and $p_{\rm max} \approx 0.3-0.8$ considered to be typical for dust in protoplanetary disks. 
However, it is unclear, how such parameters can produce the very red colors of about $\eta_{\rm V/IR}\approx +2$ for the two targets MWC~758 and LkCa~15. 
Such colors do not seem to be possible under the assumption, that the reflectivity colors are only a result of the wavelength dependence of the dust scattering parameters $\omega(\lambda)$, $g(\lambda)$ and $p_{\rm max}(\lambda)$.
Therefore, for these two objects, we consider also a scenario where the extremely red color results from a combination of a disk reflectivity with a red color and an "extra" reddening of the radiation illuminating the disk by absorbing hot dust located near the central star.

Different dust models have been tested to explain the scattered radiation of protoplanetary disks. Dust composed of large ($a\approx$ a few $\mu$m), compact aggregates are compatible with the red disk color and the estimated scattering parameters $\omega$ and $g$ \citep{Mulders2013, Min2016} but they under-predict the observed fractional polarization of the reflected radiation. 
Models for highly porous aggregates grains represent well the typical scattering parameters $\omega$, $g$ and $p_{\rm max}$, but they predict gray or blue colors for the reflected light from the disk, but not red colors \citep{Tazaki2019}. Because of these discrepancies, one should investigate alternative dust models to explain the available and future disk observations. 

Our data provide hints that the color of the disk reflectivity could be connected to the stellar irradiation in the sense that the stronger stellar irradiation in disks around Herbig stars produces redder dust reflectivity colors. 
However, we also find a trend between color and the near-IR excess from hot dust near the star, which is in our sample stronger for Herbig stars.
Because of the small sample size, both correlations are not significant and more determinations are required to clarify the presence of systematic trends between disk color and system parameters.

This study demonstrates, that we can now derive accurately the wavelength dependence from about $0.6~\mu$m to $2.2~\mu$m for the scattered radiation for many protoplanetary disks using modern near-IR and visible high-resolution imaging data obtained with adaptive optics systems \citep[e.g.,][]{Schmid2022}. 
Quantitative measurements provide now new information on the dust evolution and dust composition which are potentially very important to understand the planet formation process in these systems.

\begin{acknowledgements}
This work has been carried out within the framework of the NCCR PlanetS supported by the Swiss National Science Foundation under grants 51NF40\_182901 and 51NF40\_205606. The data are based on observations collected at the European Southern Observatory under ESO programs: 095.C-0273, 095.C-0298, 095.C-0404, 095.C-0693, 096.C-0248, 096.C-0248, 096.C-0333, 096.C-0523, 096.C-0978, 097.C-0902, 099.C-0341, 099.C-0601, 099.C-0891, 0102.C-0916, 0104.C-0472, 106.21HJ.001, 1100.C-0481, 297.C-5023, and 60.A-9358. We thank the anonymous referee for their valuable comments and insightful suggestions, which greatly contributed to the improvement of this work.

\end{acknowledgements}

\bibliographystyle{aa} 
\bibliography{biblio}

%
%
\begin{appendix}

\section{Description of the archival data}
\label{tab:obsdetails}

Table~\ref{tab:obs-info} gives details about the used archival disk observations for each target. All targets were observed at least in one visible band $\lambda_c< 1~\mu$m with SPHERE/ZIMPOL and one near-IR band with $\lambda_c> 1~\mu$m with SPHERE/IRDIS. Detailed descriptions of these instruments are given in \citet{Beuzit2019}, \citet{Dohlen2008}, and \citet{Schmid2018}.
 
\begin{figure}
    \centering
    \includegraphics[width=0.45\textwidth]{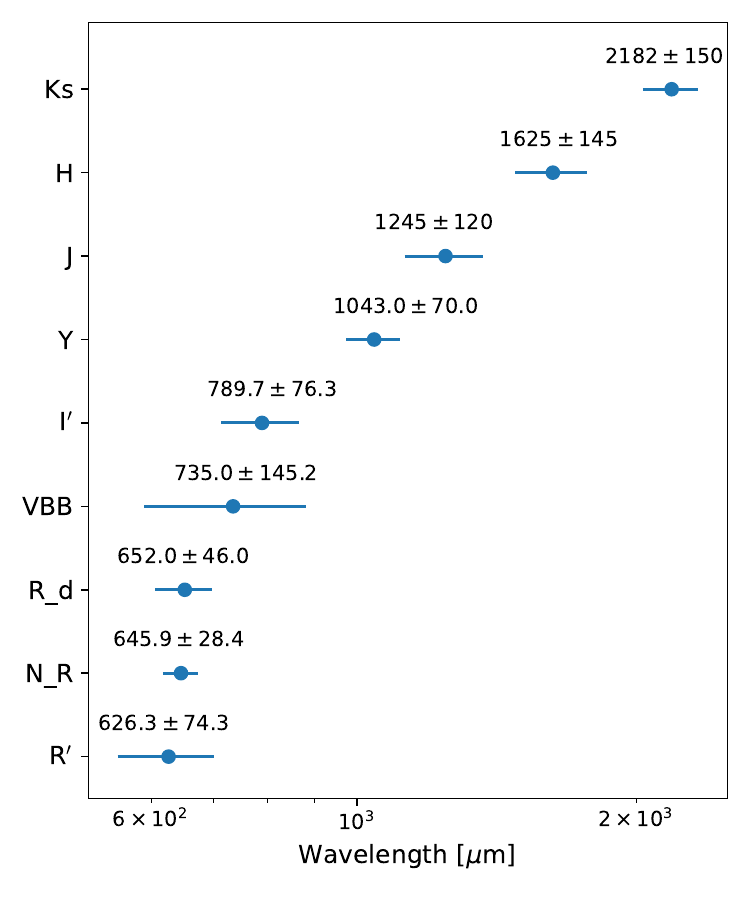}
    \caption{Central wavelengths and width at half maximum of the used filters. 
    }
    \label{fig:filter}
\end{figure}

The disks were observed in different filters and the central wavelengths $\lambda_c$ and the passband widths are given and illustrated in Fig.~\ref{fig:filter}. 
The filter band description $R'$/$I'$ indicates that the object was observed with ZIMPOL simultaneously in two different bands.
ZIMPOL has two camera arms and can therefore obtain polarimetric data for each arm using its own color filter. In most cases, the two arms use the same filter type and the result of the two arms is just averaged. The filter R$_d$ (for PDS~66, RX~J1604, and RX~J1615) describes the transmission function for the ZIMPOL instrument if the broadband R$´$ filter is combined with the dichroic beam-splitter between ZIMPOL and the wavefront sensor of SPHERE. The passband is predominantly defined by this beam-splitter. Polarimetry with IRDIS is only available in combination with the broadband filters Y, J, H, and Ks. 

Depending on the used subsystem ZIMPOL or IRDIS the meaning of some of the indicated parameters in Table~\ref{tab:obs-info} are different.
The columns "Category" and "inst. mode" describe observing configurations for the two subsystems. For ZIMPOL the instrument "Category" is always "Object", but sometimes frames with "low" counts were taken to get non-coronagraphic and unsaturated observations for the flux or PSF calibration measurements. 
This can be achieved by inserting additionally a neutral density filter, by using shorter integration times, or by setting a different gain factor for the detector read-out. All ZIMPOL observations were taken without coronagraph and in many cases, the central star is not saturated therefore no additional "low" count frames for the flux and PSF calibration are required.
ZIMPOL "inst mode" distinguishes between P1 and P2 which are both polarimetric modes but in P1 (PDS~66 data) the field rotates during the observations, while in P2 the field is stabilized. For the type of observations used in this work, the difference between P1 and P2 is only relevant for the use of the correct reduction template.

The SPHERE AO system provides IRDIS observations with a high contrast while the detector dynamic range is quite limited. For this reason, observation of protoplanetary disks in the "DPI" polarimetric mode of IRDIS are mostly taken with a coronagraph to avoid strong detector saturation at the position of the star. A typical IRDIS observation includes therefore complementary "flux" frames, taken with the star offset from the coronagraphic mask and using a neutral density filter to avoid saturation. 
No such flux frames were required for the $J$ band data of the fainter objects LkCa~15 and PDS~70, because even in the non-coronagraphic mode the central star did not saturate in the "Object" frames.

The indicated integration time nDIT $\times$ DIT applies to one out of four HWP settings of a polarimetric cycle. Thus, the total integration time of the observations given in Table~\ref{tab:obs-info} is the product of nDIT $\times$ DIT $\times$ 4 $\times$ $n_{\rm cyc}$. 

Atmospheric conditions for the observations are characterized by the atmosphere coherence time  $\tau_0$ and the seeing. Conditions can be considered to be quite good for seeing $< 0.8''$ and $\tau_0>4$~ms, but often the atmosphere is very variable, and therefore a careful correction of the PSF smearing effects is important.

\onecolumn

\begin{table*}[]
    \caption{Summary of selected observations }
    \label{tab:obs-info}
    \centering
    \resizebox{0.85\textwidth}{!}{%
    \begin{tabular}{p{2cm}p{0.9cm}p{2.2cm}p{1.2cm}p{2.2cm}p{2.1cm}p{0.6cm}p{1.4cm}p{1.6cm}}
    \hline
    \hline
    Host star  & Filters & Epochs     & Category & det mode       & nDIT $\times$ DIT    & $n_{cyc}$ & $\tau_0$[ms] & seeing[''] \\
    \hline
    RX~J1604   & R\_d & Jun 11, 2015 & Object*   & P2             & 2 $\times$ 120 s     & 6         & 2.4          & 1.44 \\
               & J       & Aug 14, 2017 & Object   & DPI + coronag. & 2 $\times$ 64 s      & 5         & 5.7          & 0.61\\
               &         &            & Flux     &  -             & 10 $\times$ 0.8375 s & 1         & 6.2          & 0.63\\
               &         & Aug 15, 2017 & Object   & DPI + coronag. & 2 $\times$ 64 s      & 4         & 3.7          & 0.91\\
               &         &            & Flux     &  -             & 10 $\times$ 0.8375 s & 1         & 3.3          & 0.77\\
               &         & Aug 18, 2017 & Object   & DPI + coronag. & 2 $\times$ 64 s      & 5         & 4.6          & 0.53\\
               &         &            & Flux     &  -             & 10 $\times$ 0.8375 s & 1         & 1.7          & 0.88\\
               &         & Aug 19, 2017 & Object   & DPI + coronag. & 2 $\times$ 64 s      & 4         & 4.0          & 0.66 \\
               &         &            & Flux     &  -             & 10 $\times$ 0.8375 s & 1         & 4.3          & 0.67\\
               &         & Aug 22, 2017 & Object   & DPI + coronag. & 2 $\times$ 64 s      & 7         & 5.9          & 0.56\\
               &         &            & Flux     & -              & 10 $\times$ 0.8375 s & 1         & 5.4          & 0.64\\
               & H       & Jul 01, 2018 & Object   & DPI + coronag. & 2 $\times$ 64 s      & 7         & 5.1          & 0.54\\
               &         &            & Flux     & ND\_1.0        & 10 $\times$ 4 s      & 1         & 3.8          & 0.80\\
    \hline
    HD~135344B & R' / I' & Apr 01, 2015 & Object   & P2             & 8 $\times$ 10 s      & 12        & 2.9-3.7      & 0.60-0.78 \\
               &         &            & Object*   & P2 + ND\_1.0   & 4 $\times$ 10 s      & 1         & 2.9-3.1          & 0.71-0.78 \\
               & J & May 03, 2016  & Object   & DPI + coronag. & 3 $\times$ 32 s      & 13        & 1.5-2.6      & 0.59-0.92\\
               &        &             & Flux     & ND\_1.0        & 2 $\times$ 0.8375 s  & 2         & 1.9-2.1      & 0.62-0.73\\
               &        & May 04, 2016  & Object   & DPI + coronag. & 2 $\times$ 32 s      & 23        & 7.0-20.8     & 0.29-1.00\\
               &        &             & Flux     & ND\_1.0        & 3 $\times$ 0.8375 s  & 2         & 9.4-12.7     & 0.54-0.78\\
               &        & May 12, 2016\tablefootmark{b}  & Object   & DPI + coronag. & 4 $\times$ 32 s      & 4         & 0.9-2.1      & 1.68-2.73 \\
               &        &             & Flux     & ND\_1.0        & 10 $\times$ 2 s      & 1         & 1.1          & 1.96  \\
               &        & Jun 22, 2016  & Object   & DPI + coronag. & 4 $\times$ 32 s      & 4         & 3.8-5.5      & 0.54-1.11\\
               &        &             & Flux     & ND\_1.0        & 4 $\times$ 0.8375 s  & 1         & 5.1          & 0.73  \\
               &        & Jun 30, 2016& Object   & DPI + coronag. & 4 $\times$ 32 s      & 4         & 5.4-9.1      &  0.29-0.50 \\
               &        &             & Flux     & ND\_1.0        & 4 $\times$ 0.8375 s  & 1         & 5.5          & 0.41  \\
    \hline
    HD~100453  & R' / I' & Mar 31, 2016\tablefootmark{b} & Object   & P2             & 8 $\times$ 10 s      & 15        & 2.4-4.6      & 0.68-1.31 \\
               &         &            & Object*   & P2 + ND\_4.0   & 4 $\times$ 10 s      & 1         & 3.6-4.1          & 0.75-0.86 \\ 
               & J       & Apr 01, 2016   & Object   & DPI + coronag.  & 4 $\times$ 32 s      & 10         &  1.6-3.5     & 0.80-1.78  \\
               &         &            & Flux     & ND\_2.0         & 5 $\times$ 4 s       & 2          &  1.8-3.1   & 0.90-1.59  \\
    \hline
    LkCa15     & VBB     & Feb 02, 2015 & Object   & P2              & 8 $\times $ 20 s     & 7        &  14.3-27.9    & 0.39-0.80 \\
               &         &            & Object*     & ND\_1.0         & 8 $\times $ 20 s     & 1        &  32.7         & 0.36     \\
               &         & Feb 12, 2015 & Object   & P2              & 8 $\times $ 20 s     & 6\tablefootmark{a}&  4.6-8.4      & 0.61-1.21 \\
               &         &            & Object*   & ND\_1.0         & 8 $\times $ 20 s     & 1        &  5.2          & 1.04 \\
               
               & J       & Dec 19, 2015\tablefootmark{c} & Object   & DPI             & 40 $\times$ 0.8375 s & 3        & 2.6-3.1       & 0.59-0.69\\
               & Ks      & Dec 08, 2020 & Object   & coronag.        & 2 $\times$ 16s       & 35       & 7.9-13.9      & 0.35-0.53\\
               &         &            & Flux     & ND\_1.0         & 5 $\times$ 2s        & 2        & 9.3           & 0.39-0.48\\      
    \hline
    TW Hydrae  & R' / I' & Apr 01, 2015 & Object*   & P2   & 4 $\times$ 10 s      & 4\tablefootmark{a}&  2.6-4.7   & 0.52-0.93 \\
               & H       & Apr 01, 2015 & Object   & DPI + coronag.  & 4 $\times$ 16 s      & 23\tablefootmark{a}  & 2.7-4.3    &  0.56-0.89\\
               &         &          & Flux     & ND\_1.0         & 1 $\times$ 0.8375 s  & 3       & 2.5-2.9    & 0.83-0.89\\
    \hline
    PDS~66     & N\_R   & Mar 19, 2016 & Object & P1 & 6$\times$ 48 s & 2 & 1.1-2.2 & 0.81-1.24 \\
               &        &            & Object*   &    & 6$\times$ 24 s & 1 & 1.9     & 0.95      \\
               & J      & Mar 15, 2016 & Object   & DPI + coronag.  & 2 $\times$ 64 s      & 6       &  1.9-3.0 & 0.96-1.52 \\
               &        &            & Flux     & ND\_1.0         & 2 $\times$ 4 s       & 1       & 1.9    & 1.5\\
               & H      & Mar 16, 2016 & Object   & DPI + coronag.  & 2 $\times$ 64 s      & 4       & 2.2-3.3     & 0.79-1.15\\
               &        &            & Flux     & ND\_1.0         & 2 $\times$ 2 s       & 2       &  2.9     & 0.90-0.92 \\
    \hline
    PDS~70     & VBB    & Jul 09, 2015 & Object*  & P2             & 6 $\times$ 40 s       & 10\tablefootmark{a}  & 1.1-2.0 & 0.81-1.66 \\
               & J     & Aug 01, 2017 & Object  & DPI            & 20 $\times$ 2 s       & 7      & 2.0-3.2 & 0.56-0.77 \\
               & H     & Aug 08, 2019 & Object  & DPI + coronag. & 1 $\times$ 64 s       & 9      & 2.5-3.4 & 0.39-0.64 \\
               &        &            & Flux    & ND\_1.0        & 3 $\times$ 4 s        & 1      & 2.5     & 0.62      \\
               & Ks    & Jul 12, 2019 & Object  & DPI + coronag. & 1 $\times$ 64 s       & 20\tablefootmark{a}    & 2.8-5.4 & 0.37-0.79 \\
               &        &            & Flux    & ND\_1.0        & 10 $\times$ 2 s       & 3      & 3.5-4.6 & 0.45-0.60 \\
    \hline
    
    RX~J1615   & R\_d   & Jun 10, 2015 & Object*  & P2  & 2 $\times$ 120 s      & 6      & 2.0-2.6 & 1.19-1.61 \\
               & J      & Jun 06, 2015 & Object & DPI + coronag. & 3 $\times$ 64 s & 6 & 1.2-1.8 & 1.65-2.45\\
               &     &             & Flux   & - & 4 $\times$ 0.8375 s & 2 & 1.8 & 1.61 \\
               & H    & Mar 15, 2016 & Object & DPI + coronag. & 2 $\times$ 64 s & 10\tablefootmark{a} & 1.5-3.6 & 0.80-2.17\\
               &      &            & Flux   & ND\_1.0       & 2 $\times$ 4 s  & 1   & 1.9    & 1.53 \\
               
    \hline
    MWC~758    & VBB    & Dec 18, 2015 & Object*  & P2             & 10 $\times$ 4 s       & 15     & 1.7-2.7 & 0.63-0.98 \\
               & B\_Y   & Nov 18, 2019 & Object  & DPI + coronag. & 2 $\times$ 32 s       & 10     & 3.9-7.0 & 0.53-0.90\\
               &        &            & Flux    & ND\_1.0        & 5 $\times$ 0.8375 s   & 1      & 5.1     & 0.64    \\ 
    \hline
    HD~169142  & R'/I'  & Jul 09, 2015 & Object* & P1 & 4 $\times$ 6s\tablefootmark{d} & 16 & 1.3-1.8 & 0.90-1.23\\
               & J      & May 03, 2015  & Object & DPI + coronag. & 5 $\times$ 32 s       & 5     & 1.7-2.8 & 0.46-0.76\\
               &        &              & Flux   & ND\_1.0        & 2 $\times$ 0.8375 s   & 1     & 2.4     & 0.55 \\
    \hline
    HD~142527  & VBB    & Mar 31, 2016  & Object* & P2             & 18 $\times$ 2s       & 12    & 3.2-5.0 & 0.61-0.98 \\
               & H      & Jun 01, 2017 & Object & DPI + coronag. & 12 $\times$ 8s        & 9     & 9.1-15.2 & 0.54-0.92\\
               &        &               & Flux  & ND\_2.0         & 8 $\times$ 2s        & 4     & 10.3-14.8 & 0.56-0.80\\
    \hline 
    \end{tabular}}
    \tablefoot{\tablefoottext{*}{Object frames with the center unsaturated and used as stellar flux calibration.}
    \tablefoottext{a}{Part of the cycles are selected due to centering failure. The four best cycles were selected for TWHya due to the low-wind effect.}
        \tablefoottext{b}{Cloudy weather conditions according to the Astronomical site monitor (ASM).}
        \tablefoottext{c}{No record in the ASM.}
        \tablefoottext{d}{First four cycles are taken with 6 $\times$ 6s.}
        }
\end{table*}
\twocolumn
\normalsize

\section{Tests of the PSF convolution correction}
In this section, we investigate the precision of the PSF convolution correction procedure for the determination of the intrinsic polarized intensity $\hat{Q}_\varphi$ for the reflected light of the disk.
For this, we simulate the convolution effects for parametric test disk models $\bar{Q}_\varphi$ and calculate the polarization of the convolved disk $\tilde{Q}_\varphi$, which represents the observed value $Q_\varphi$. We then apply the correction map method $\hat{Q}_{\varphi}(x, y) = \Tilde{Q}_{\varphi}(x, y)\times f(x,y)$ as described in Sect.~\ref{sect:corr-map} and compare corresponding integrated polarized intensities for the corrected value $\hat{Q}_\varphi$ with the test model value $\bar{Q}_\varphi$. 
For the disk models, we adopt a prescription for an inclined disk with an intrinsically axisymmetric geometry (an inclined ring), with an intensity contrast between the front side and the back side to simulate a strong azimuthal asymmetry for the polarized scattered light. 

\begin{figure}
    \centering
    \includegraphics[width=0.45\textwidth]{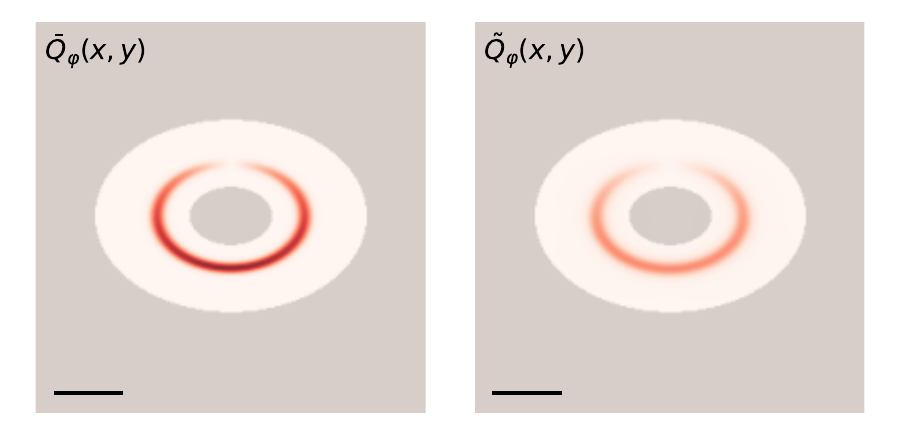}
    \caption{Example of an inclined ring-shaped disk test model $\bar{Q}_{\varphi}(x, y)$. The model assumes $r_0=0.55 \arcsec$, $\sigma =0.04\arcsec$, and inclination $45^\circ$. $\tilde{Q}_{\varphi}$ is obtained by convolving $\bar{Q}_{\varphi}(x, y)$ with the $J$ band "good" PSF}
    \label{fig:asy-image}
\end{figure}

The brightness distribution for the polarization of these disks is given by 
\begin{align}
    \bar{Q}_{\varphi, r_0}(r, \varphi) &=  \exp \Big(-\dfrac{(r-r_0)^2}{2\sigma^2}\Big)|\sin\frac{\varphi}{2}|\,,\\
    r^2 &= x^2 + (y/\cos i)^2\,,
\end{align}
where $\varphi$ is the angle with respect to the minor axis of the disk back side. 
Figure~\ref{fig:asy-image} illustrates the disk structure for an inclination of $i=45^\circ$, $r_0=0.55''$ and $\sigma = 0.04''$. The sine function yields zero flux for the backside $Q_{\varphi, r_0}(r, \varphi=0) = 0$ and this exaggerates the front-to-back side asymmetry when compared to the observed disk, but it provides a good test case for the investigation of the impact of strong variation of the azimuthal brightness distribution of disks on the PSF correction. 

\begin{figure}
    \centering
    \begin{subfigure}[]{0.45\textwidth}
        \includegraphics[width=\textwidth]{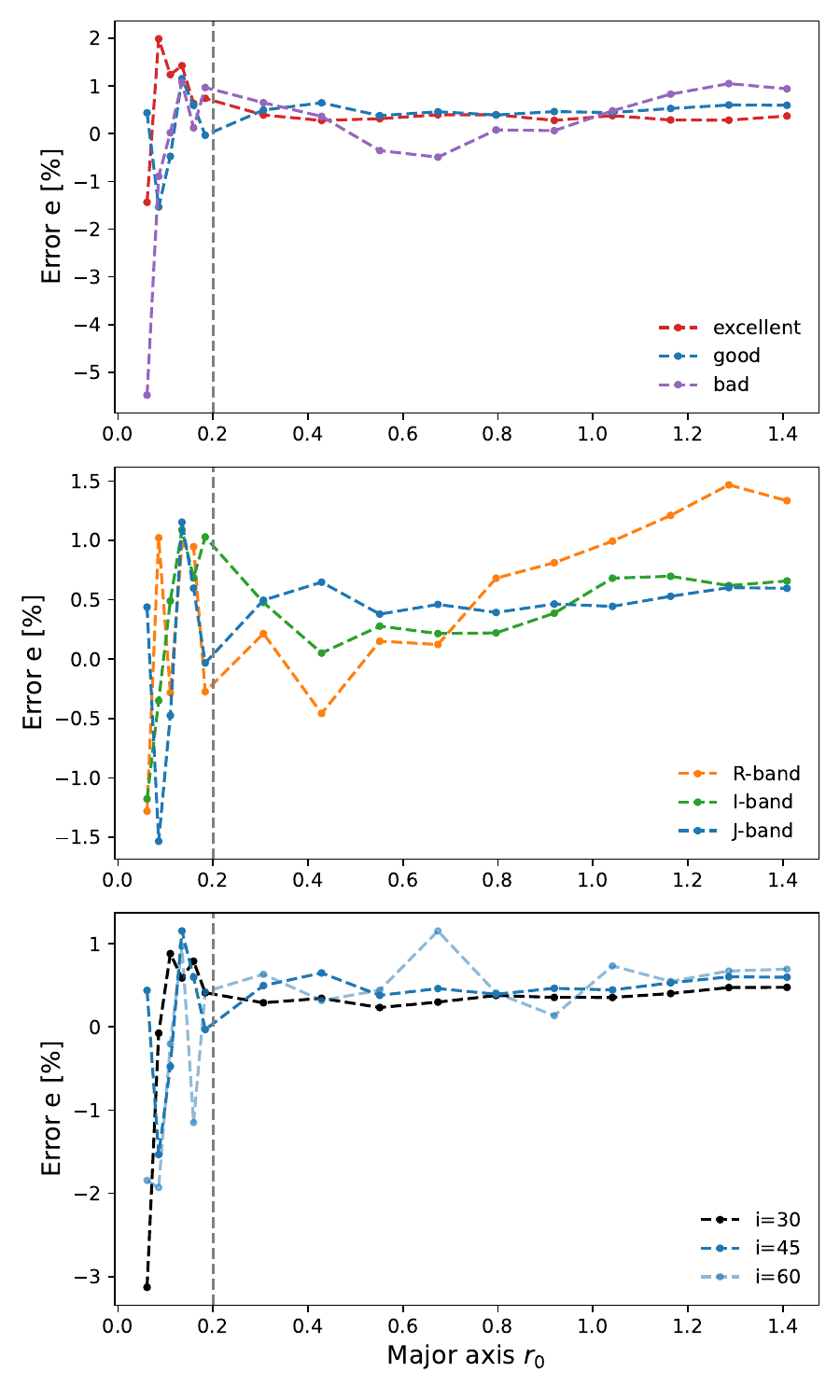}
    \end{subfigure}
    \caption{Relative error $e=(\hat{Q}_\varphi-\bar{Q}_\varphi)/\bar{Q}_\varphi$ between the corrected disk intensity $\hat{Q}_\varphi$ and test model intensity $\bar{Q}_{\varphi}$ for inclined Gaussian rings. 
    Top: For $J$ band PSFs taken under excellent, good, and bad conditions for $i=30^\circ$; 
    Middle: For $R'$, $I'$, and $J$ band PSFs taken under good conditions for $i=30^\circ$; Bottom: For $J$ band PSF taken under good conditions but different $i=30^\circ$, $45^\circ$, and $60^\circ$.}
    \label{fig:err-incline}
\end{figure}

We calculate the relative difference or error between the corrected values and the model values $e=(\hat{Q}_\varphi-\bar{Q}_\varphi)/\bar{Q}_\varphi$. 
Models with different disk parameters and convolutions with different PSF were investigated. Errors $e$ are calculated and the results are presented in Fig.~\ref{fig:err-incline} as function of apparent disk radius $r_0$ for disks with an inclination $i=30^\circ$ using the $J$ band PSFs classified in Table~\ref{tab:psf_quality} as excellent, good, and bad, and the "good" PSFs for the $R'$, $I'$ and $J$ bands.
For the good PSF in the $J$ band also different inclinations $i=30, 45, $ and $60^\circ$ are investigated. 

The results show that the correction map procedure produces for disks larger than $r_0 > 0.2''$ typically very small errors $e$ or relative differences $<1~\%$ between corrected polarized disk intensities after convolution $\hat{Q}_\varphi$ and test model intensities $\bar{Q}_\varphi$ with a tendency that $\hat{Q}_\varphi$ are slightly higher than $\bar{Q}_\varphi$. The errors are particularly small for higher quality PSFs and lower disk inclination. 
For small disks $r\leq 0.2''$ the errors are a bit larger, at a level of a few percent, with a tendency that the corrected values $\hat{Q}_\varphi$ are slightly lower than the test model values $\bar{Q}_\varphi$.
In our disk sample, there are only HD~169142 with a small disk ring $r<0.2\arcsec$. For simplicity reasons, we adopt $2\%$ as for the systematic uncertainty introduced by the simulation of the PSF convolution correction. 

\end{appendix}

\end{document}